\documentclass{emulateapj}
\usepackage{graphicx}

\def\etal{et al.~}

\def\ba{\begin{eqnarray}}
\def\ea{\end{eqnarray}}

\begin{document}

\title{Planet formation in stellar binaries II: 
overcoming the fragmentation barrier in $\alpha$ Centauri and 
$\gamma$ Cephei-like systems}

\author{Roman R. Rafikov\altaffilmark{1} \& Kedron Silsbee\altaffilmark{1}}
\altaffiltext{1}{Department of Astrophysical Sciences,
Princeton University, Ivy Lane, Princeton, NJ 08540; 
rrr@astro.princeton.edu}


\begin{abstract}
Planet formation in small-separation ($\sim 20$ AU) eccentric 
binaries such as $\gamma$ Cephei or $\alpha$ Centauri is believed to be 
adversely affected by the presence of the stellar companion. Strong 
dynamical excitation of planetesimals by the eccentric companion can result 
in collisional destruction (rather than growth) of $1-100$ km objects, 
giving rise to the ``fragmentation barrier'' for planet formation.
We revise this issue using a novel description of secular dynamics 
of planetesimals in binaries, which accounts for the gravity 
of the eccentric, coplanar protoplanetary disk, as well as gas drag. 
By studying planetesimal collision outcomes we show, in contrast 
to many previous studies, that planetesimal growth and subsequent 
formation of planets (including gas giants) in AU-scale orbits  
within $\sim 20$ AU separation binaries may be possible, provided that 
the protoplanetary disks are massive ($\gtrsim 10^{-2}M_\odot$) 
and only weakly eccentric (disk eccentricity $\lesssim 0.01$). 
These requirements are compatible with both the existence of massive
(several $M_J$) planets in $\gamma$ Cep-like systems and the results 
of recent simulations of gaseous disks in eccentric binaries.
Terrestrial and Neptune-like planets can 
also form in lower-mass disks at small (sub-AU) radii. 
We find that fragmentation barrier is less of a problem in 
eccentric disks which are apsidally aligned with the binary orbit.
Alignment gives rise to special locations, where 
(1) relative planetesimal velocities are low and (2) the timescale 
of their drag-induced radial drift is long. This causes
planetesimal pileup at such locations in the disk and promotes their 
growth.
\end{abstract}




\section{Introduction.}  
\label{sect:intro}


Planets are known to be able to form in a variety of 
environments, some of which are believed to be hostile 
to their genesis. A good illustration of this statement is 
provided by planets detected in close binary systems, such 
as $\gamma$ Cephei (Hatzes \etal 2003). This eccentric 
($e_b=0.41$), relatively small semi-major axis ($a_b=19$ AU) system 
consists of two stars of mass $M_p=1.6M_\odot$ and 
$M_s=0.41M_\odot$. It harbors a giant planet with the projected 
mass $M_{\rm pl}\sin i=1.6M_J$ in orbit
with semi-major axis $a_{\rm pl}\approx 2$ AU and eccentricity 
$e_{\rm pl}\approx 0.12$ around the primary. 

Several more such systems of S-type in 
classification of Dvorak (1982) are known at present (Chauvin \etal 
2011; Dumusque \etal 2012). Two of them --- HD 196885 
(Correia \etal 2008) and HD 41004 (Zucker \etal 2004) 
harbor giant planets in orbits with $a_{pl}=1.6-2.6$ 
AU. Two more --- $\alpha$ Cen (Dumusque \etal 2012) and 
Gl 86 (Queloz \etal 2000) host planets at smaller separations, 
$a_{\rm pl}\approx 0.04$ AU and $a_{\rm pl}\approx 0.11$ AU, 
correspondingly. These systems exhibit a diversity of 
planetary masses, with an Earth-like planet 
($M_{\rm pl}\sin i=1.1M_\oplus$) orbiting our neighbor $\alpha$ 
Cen (cf. Hatzes 2013), and other binaries hosting gas giants with 
$M_{\rm pl}\sin i=(1.6-4.0)M_J$ (Chauvin \etal 2011).

The existence of planets in these tight binaries has been a 
serious challenge for planet formation theories. The expectation 
of inner rather than outward planet migration due to 
disk-planet interaction (Ward 1986) suggests that such planets 
form in situ, at $1-2$ AU (as we show in this work 
it is very difficult to form them even further out). At these 
separations gravitational instability is a very unlikely avenue 
of planet
formation (Rafikov 2005, 2007). An alternative model of core 
accretion (Harris 1978; Mizuno 1980) relies on formation of a 
massive core by 
collisional agglomeration of a large number of planetesimals, 
possibly starting at small, $\lesssim 1$ km, sizes. It is this 
stage of planetesimal growth in tight binaries that presents 
significant problems to existing planet formation theories.

Indeed, it has been known since the work of Heppenheimer (1978)
that an eccentric stellar companion can drive very large 
planetesimal eccentricities, $\sim 0.1$ at AU-scale separations.
This would cause planetesimals to collide at high relative 
speeds of a few km s$^{-1}$. As this is much higher 
than the escape speed from the surface of even a $100$ km object 
(about 100 m s$^{-1}$), collisions between planetesimals should 
lead to their {\it destruction} rather than growth, introducing 
a {\it fragmentation barrier} for planet formation 
(see \S \ref{sect:coll}). The 
theoretical expectation of suppressed planet formation in 
$a_b<20$ AU binaries has been largely corroborated by 
recent observations (Wang \etal 2014). 

The premise of our present work is that the key to solving 
the fragmentation barrier puzzle lies in better understanding of 
planetesimal 
dynamics. However, some alternative suggestions have also been 
considered over the years. For example, Th\'ebault \etal (2008, 2009) 
proposed that tight planet-hosting binaries could have started 
on more extended orbits, which were subsequently shrunk 
by interactions with other stars in their birth cluster.
Paardekooper \& Leinhardt (2010) propose a solution involving
a non-standard mode of planetesimal accretion. It may also 
be possible that planetesimals are born big (Johansen \etal 2007), 
with sizes exceeding $10^2$ km, in which case they are safe from
collisional destruction from the start. These possibilities would 
need to be 
invoked if we were not able to resolve the fragmentation barrier 
puzzle by the better treatment of planetesimal dynamics alone, 
underscoring the importance of this aspect of the problem.

Dynamics of planetesimals in binaries are complicated by a 
plethora of agents affecting their motion. It has been long 
realized that both the companion gravity and gas drag affect 
planetesimal motion (Marzari \& Scholl 2000; Th\'ebault 
\etal 2004). However, subsequently it has also been understood 
that these processes alone cannot overcome the fragmentation 
barrier (Th\'ebault \etal 2008). More recently, it was shown 
that the gravity of protoplanetary disk in which planetesimals 
reside has a dominant effect on their dynamics (Rafikov 2013b; 
hereafter R13). 
The tendency of protoplanetary disks in binaries to become 
eccentric further complicates this issue, see Silsbee \& Rafikov
(2013; hereafter SR13).  

Generalizing these efforts, 
Rafikov \& Silsbee (2014; hereafter Paper I) combined different physical
ingredients --- gravity of an eccentric disk, perturbations due 
to the companion star, and gas drag --- to present a unified 
picture of planetesimal dynamics in binaries in {\it secular} 
approximation. They came up with analytical 
solutions for planetesimal eccentricity, and explored the behavior 
of relative velocities between planetesimals of different sizes.

Our present goal is to use these dynamical results to
understand planetesimal growth in tight binaries with particular
focus on the fragmentation barrier issue. 
We couple them with recent understanding of collisional 
fragmentation based on the work of Stewart \& Leinhardt (2009) 
and explore the conditions under which planetesimals can grow 
unimpeded by fragmentation in situ, i.e. at the present day
orbits of planets in tight binaries.
We do this for a variety of different collisional criteria 
governing planetesimal growth and carefully explore the space
of various disk+binary parameters.  
To summarize our main finding from the start, we find that 
even in tight binaries planet formation should be possible 
in {\it massive} protoplanetary disks which are {\it only 
weakly eccentric}.

This paper is structured as follows. We summarize the main
dynamical results of Paper I in \S \ref{sect:dyn_sum}. We 
describe our treatment of planetesimal collision outcomes in  
\S \ref{sect:coll}. Conditions for planetesimal growth in 
non-precessing and precessing disks are determined in 
\S \ref{sect:pl_form} and \S \ref{sect:pl_prec}
correspondingly. Sensitivity of our results to model
parameters is explored in \S \ref{sect:sens}. Radial 
migration of planetesimals is covered in 
\S \ref{sect:pl_migr}. Implications of our results for planet 
formation can be found in \S \ref{sect:bin_pl_form}.
We summarize our main conclusions in \S   
\ref{sect:summ}.


\section{General setup.}  
\label{sect:setup}


We study planet formation in binaries using a setup similar 
to SR13 and Paper I. The binary with semi-major axis $a_b$ 
and eccentricity $e_b$ has components with masses $M_p$ 
(primary) and $M_s$ (secondary). We define $\nu\equiv M_s/M_p$. 
The primary star is orbited by an eccentric protoplanetary 
disk, coplanar with the binary orbit. Fluid elements in the 
disk follow elliptical trajectories with the primary star 
in the focus. We adopt a power law dependence of the gas 
eccentricity $e_g(a_d)$ as a function of the semi-major 
axis $a_d$ of a particular ellipse:
\ba
e_g(a_d)=e_0\left(\frac{a_{\rm out}}{a_d}\right)^q.
\label{eq:e0}
\ea
Here $a_{\rm out}$ is the outer cutoff radius of the disk. 
Simulations show that in eccentric binaries 
with $e_b=0.4$, the disk gets truncated at 
$a_{\rm out}\approx (0.2-0.3)a_b$ by gravitational 
perturbations from the companion. Thus, $e_0$ is the 
eccentricity of fluid trajectories 
at the outer edge of the disk, $a_d=a_{\rm out}$.

For simplicity all fluid trajectories are assumed to have 
{\it aligned} apsidal lines, so that the disk orientation is 
uniquely defined via a single parameter $\varpi_d$ --- the angle 
between the disk and binary apsidal lines. 

We let $\Sigma_p(a_d)$ be the disk surface density at the periastron 
of the fluid trajectory with semi-major axis $a_d$. Surface density 
at an arbitrary point in the disk can be uniquely specified once 
$e_g(a_d)$ and $\Sigma_p(a_d)$ are known (Statler 2001; Ogilvie 2001; 
SR13). Here we assume a power law dependence of $\Sigma_p$ between 
$a_d=0$ and $a_{\rm out}$. Assuming that disk contains mass $M_d$ out to 
$a_{\rm out}$ the surface density distribution is given by  
\ba
\Sigma_p(a_d) & = & \frac{2-p}{2\pi}\frac{M_d}{a_{\rm out}^2}
\left(\frac{a_{\rm out}}{a_d}\right)^p
\label{eq:sig_0}\\
& \approx & 3\times 10^3~
\mbox{g cm}^{-2}M_{d,-2}a_{\rm out,5}^{-1}a_{d,1}^{-1},
\nonumber
\ea
where $p$ is the power law index ($p=1$ in the numerical estimate), 
$M_{d,-2}\equiv M_d/(10^{-2}M_\odot)$, $a_{\rm out,5}\equiv a_{\rm out}/(5$ 
AU) and $a_{d,1}\equiv a_d/$AU. Equation (\ref{eq:sig_0}) 
neglects disk ellipticity and assumes $p<2$, so that most 
of the disk mass is concentrated near $a_{\rm out}$.
Unless stated otherwise (see \S \ref{sect:pl_form}) we will 
be using a disk model with $p=1$ and $q=-1$ in our 
calculations, i.e. $\Sigma_p(a_d)\propto a_d^{-1}$ and 
$e_g(a_d)\propto a_d$; see R13 and SR13 for 
motivation. We assume a disk with $a_{\rm out}$ = 5 AU.

Planetesimals of radius $d_p$ orbit the primary within 
the disk and are coplanar 
with it and the binary. Their orbits are described by semi-major 
axis $a_p$, eccentricity $e_p$ and the apsidal angle (w.r.t. 
the binary apsidal line) $\varpi_p$. The latter two are
often combined for convenience into the planetesimal 
eccentricity vector ${\bf e}_p=(k_p,h_p)=
e_p(\cos\varpi_p,\sin\varpi_p)$. Everywhere in this work we 
assume $e_p\ll 1$ as well as $e_g\ll 1$.


\section{Summary of the results on planetesimal dynamics.}  
\label{sect:dyn_sum}


In Paper I we obtained a number of important results on
the dynamics of planetesimals in binaries in secular 
approximation, i.e. neglecting short-term gravitational 
perturbations (Murray \& Dermott 1999). Our calculations 
simultaneously accounted for the gravity of the massive 
eccentric protoplanetary disk, binary companion, and gas 
drag. 

Gravitational perturbations due to the binary companion and  
the eccentric disk excite planetesimal eccentricity  
at the rates determined by the eccentricity excitation 
terms $B_b$ due to binary and $B_d$ due to disk,
given by equations (7,PI) and (8,PI), correspondingly 
(``PI'' means that the referenced equation can be found in 
Paper I). At the same time, the axisymmetric component of 
the gravity of these perturbers drives apsidal precession of 
planetesimal orbits at rates $A_b$ (binary, equation (5,PI)) 
and $A_d$ (disk, equation (6,PI)). We invariably find 
that in disks massive enough to form Jupiter mass planets,
$M_d\gtrsim 10^{-2}M_\odot$, planetesimal precession, and 
often eccentricity excitation, are dominated out to a few 
AU by the gravity of the disk. This finding is a novel 
result of R13, SR13 and Paper I. 

We showed that in the case of a non-precessing disk with a 
fixed orientation with respect to the binary apsidal line 
planetesimal eccentricity ${\bf e}_p$ is an analytic function
of the planetesimal size $d$ and system parameters,
given by the expressions (22,PI)-(28,PI), (32,PI), \& (33,PI). 
The latter enter equations through the two key variables --- 
characteristic eccentricity $e_c$ and size $d_c$, defined by 
equations (29,PI) and (31,PI), correspondingly. Dependence 
of $e_c$ and $d_c$ on the system parameters was explored in 
great detail. 

\begin{figure}[t]
\vspace{-1cm}
\epsscale{1.2}
\plotone{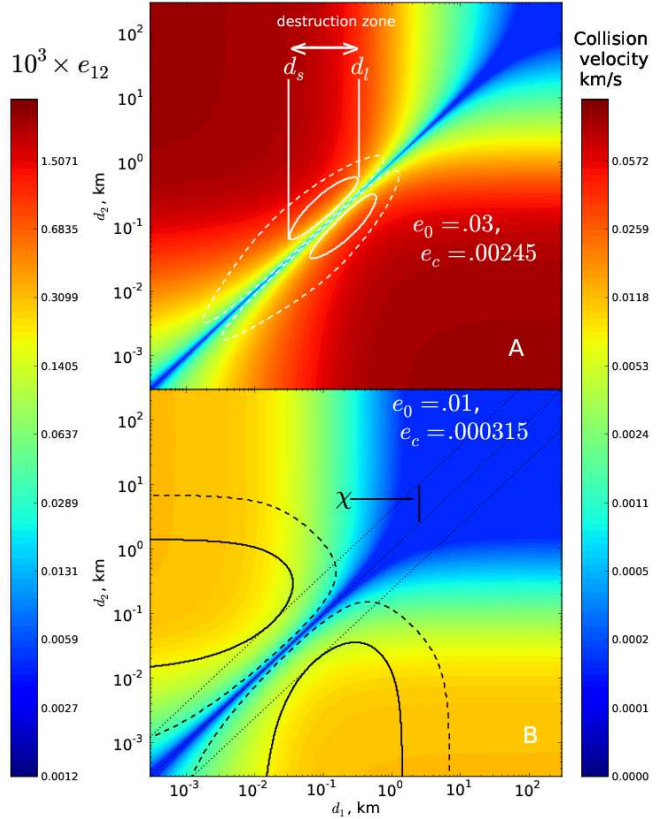}
\caption{
Maps of the relative eccentricity $e_{12}$ (left color bar)
and velocity $v_{12} = e_{12} v_k$ (right color bar) for planetesimals of 
different sizes $d_1$ and $d_2$ (see Paper I for similar maps). 
Calculation is done at $a_p=1$ AU for $\gamma$ Cephei system 
for our standard ($p=1,q=-1$), aligned ($\varpi_d=0$) disk 
with $M_d/M_p = 0.01$ and disk eccentricity at its outer edge 
(a) $e_0=0.03$ (resulting in $e_c\approx 2.45\times 10^{-3}$) and 
(b) $e_0=0.01$ ($e_c\approx 3.15\times 10^{-4}$). 
Contours illustrate collisional outcomes using 
different fragmentation criteria: catastrophic destruction 
(\ref{eq:coll_prescr}) in panel (a) --- white, and erosion 
(\ref{eq:erosion}) in panel (b) --- black. Planetesimals are destroyed 
in collisions of pairs of objects within corresponding 
contours. Solid and dashed contours are for strong and weak 
planetesimals. The extent of the destruction zone (arrow) and 
the smallest and largest ($d_s$ and $d_l$) sizes of planetesimals 
that get destroyed are illustrated in panel (a). In panel (b) 
parameter $\chi$ measures the extent of the erosion zone:
it represents a lower limit on the size ratio of objects 
that lead to erosive collisions.
\label{fig:coll_outcomes}}
\end{figure}

Our analytic solutions allow us to produce maps of the
relative eccentricity $e_{12}=|{\bf e}(d_1)-{\bf e}(d_2)|$ 
for pairs of planetesimals of different sizes $d_1$ and $d_2$;
an example is shown in Figure \ref{fig:coll_outcomes}. 
We also derived a distribution of approach velocities for 
colliding planetesimals (\S 8 of Paper I) and shown it to be 
rather narrow, with the approach velocity $v_{12}$ constrained to lie 
within the range $(1/2)v_K e_{12}<v_{12}<v_K e_{12}$, where 
$v_K$ is the local Keplerian speed. Thus, maps such as 
shown in Figure \ref{fig:coll_outcomes} directly characterize 
the typical velocity at which planetesimals collide, 
$v_{12}\sim v_K e_{12}$, and allow us to understand their 
collision outcomes, see \S \ref{sect:coll}.

The $e_{12}$ and $v_{12}$ maps in Figure \ref{fig:coll_outcomes} 
are made for $\gamma$ Cephei system at $a_p=1$ AU for the 
standard ($p=1$, $q=-1$) aligned ($\varpi_d=0$) disk 
with $M_d/M_p = 10^{-2}$ and 
$e_0=0.03,0.01$ (resulting in $e_c=2.45\times 10^{-3},
3.15\times 10^{-4}$ correspondingly). One can clearly see that 
planetesimals exhibit small relative eccentricity in
a blue region around the diagonal line $d_1=d_2$. 
This low-$e_{12}$ ``valley'' appears because planetesimals with 
similar sizes follow similar orbits, and collide with low 
relative speed. The valley is narrowest at $d_{1},d_{2}\sim 0.1-1$
km (depending on $e_c$), which corresponds to the characteristic 
size $d_c$ given by equation (31,PI). For $d_{1},d_{2}\ll d_c$
planetesimals experience apsidal alignment and their 
relative eccentricities are lowered by gas drag. For
$d_{1},d_{2}\gg d_c$, apsidal alignment is accomplished by the
disk and companion gravity, again resulting in small 
$e_{12}$. On the contrary, planetesimals of very different 
sizes (upper left and lower right regions) are not aligned 
and exhibit high relative eccentricity, with $e_{12}\approx e_c$ 
given by equation (29,PI). 

We also obtained some analytical results on planetesimal
eccentricity behavior in {\it precessing} disks, see \S 6 of 
Paper I. We did this in two limiting cases: when binary
gravity dominates over that of the disk, and vice versa. 
These asymptotic results are used to understand planetesimal
growth in precessing disks in \S \ref{sect:pl_prec}.


\section{Planetesimal collision outcomes.}  
\label{sect:coll}


Description of the dynamical behavior of planetesimals provided
in Paper I is used in this work to understand the outcomes 
of their collisions. 

There are different ways in which planetesimal collisional evolution 
can be characterized. A high-velocity collision is usually considered 
{\it catastrophic} when the mass of the largest surviving remnant 
is less than half of the combined mass of objects 
$M_{\rm{tot}}=m_1+m_2$ involved. In this work we use a fragmentation 
prescription developed by Stewart \& Leinhardt (2009), which
suggests that a collision is catastrophically disruptive if 
\ba
&& \frac{Q_{\rm R}}{Q^*_{\rm{RD}}}>1,
\label{eq:coll_prescr}\\
&& Q_{\rm R}=\frac{M_{\rm r} v_{\rm{coll}}^2}{2M_{\rm{tot}}},
\label{eq:QR}\\
&& Q^*_{\rm{RD}}=
q_s R_{C1}^{9\mu_c/(3-2\phi)}v_{\rm{coll}}^{2-3\mu_c} + 
q_gR_{C1}^{3\mu_c}v_{\rm{coll}}^{2-3\mu_c},
\label{eq:QRstar}
\ea
where $Q_{\rm R}$ is the appropriately scaled kinetic energy 
of the collision, $M_{\rm r}=m_1 m_2/(m_1+m_2)$ is the reduced 
mass of the colliding objects, and $v_{\rm coll}$ is the collision
speed at the moment of contact. 
The energy threshold for catastrophic disruption $Q^*_{\rm{RD}}$ 
depends on constants $q_s$, $\mu_c$, $\phi$, and $q_g$ related to 
the material properties of the planetesimals; $R_{C1}$ is the 
radius of a sphere with the mass $M_{\rm{tot}}$ and a density 
of 1 g cm$^{-3}$. Following Stewart \& Leinhardt (2009), we use 
$\mu_c = 0.4$, $\phi =7$, $q_s = 500$, and $q_g = 10^{-4}$ 
(in proper CGS units) for our weak planetesimals and
$\mu_c = 0.5$, $\phi =8$, $q_s = 7\times 10^4$, and 
$q_g = 10^{-4}$ for strong ones.

On the other hand, even if the condition (\ref{eq:coll_prescr})
is not satisfied and catastrophic disruption is avoided, collisional
growth is not guaranteed --- it requires that the largest object
(e.g. $m_1$) is not {\it eroded} in a collision. Erosion occurs
when the largest remnant is less massive than the more massive 
body involved in a collision. According to Stewart \& Leinhardt 
(2009) erosion happens whenever
\ba
\frac{Q_{\rm R}}{Q^*_{\rm{RD}}}>2\frac{m_2}{M_{\rm tot}},~~~~m_2<m_1.
\label{eq:erosion}
\ea
This condition is far more prohibitive for growth than 
(\ref{eq:coll_prescr}) since
$m_2$ can be much less than $m_1$. Growth in a given collision 
occurs only when the condition (\ref{eq:erosion}) is violated.

\begin{figure}[h]
\epsscale{1.2}
\plotone{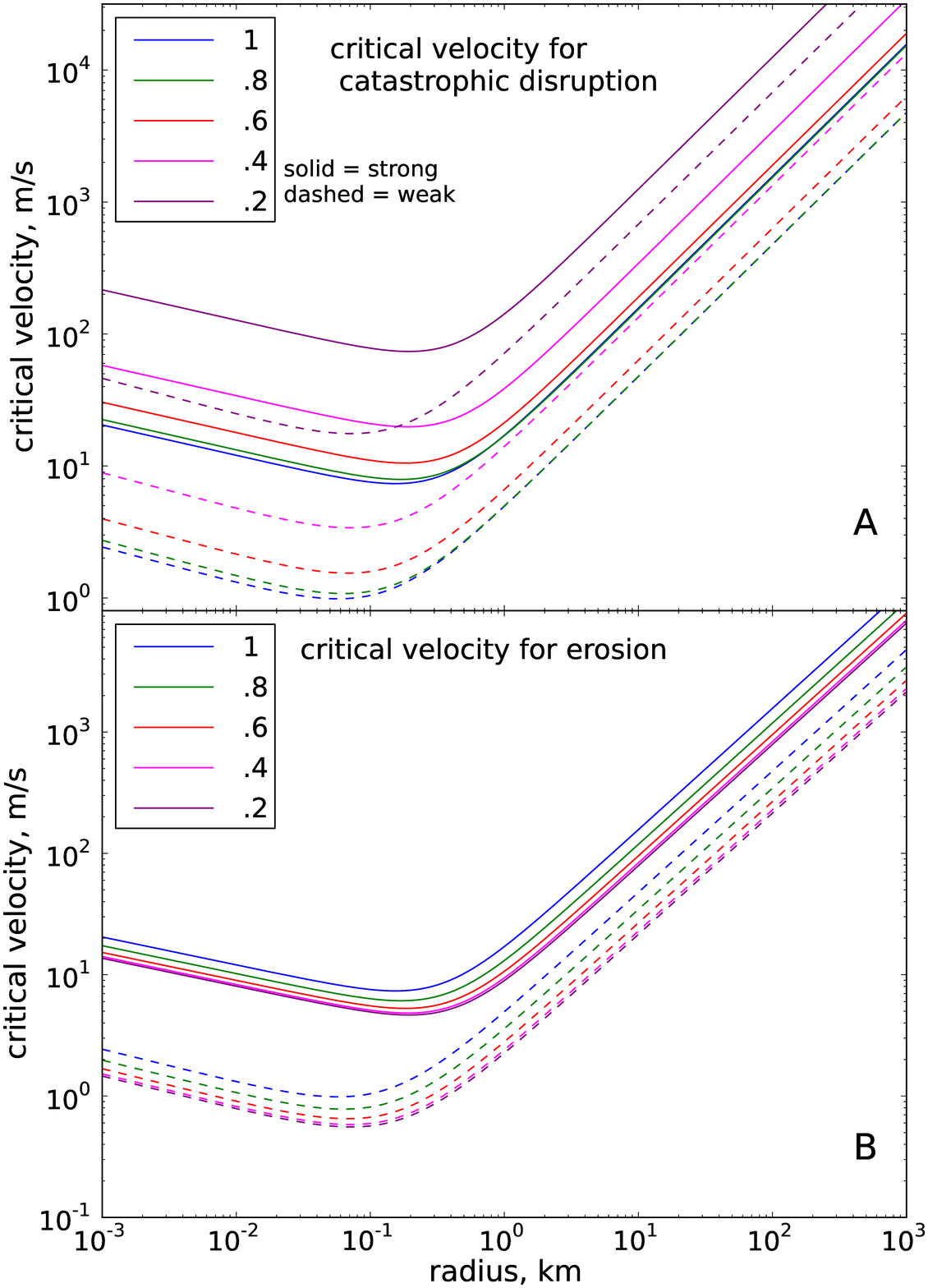}
\caption{
Critical velocity curves, shown as a function of the 
bigger planetesimal radius, for different collisional conditions
proposed in Stewart \& Leinhardt (2009) and used in this work: 
(a) catastrophic disruption, equation (\ref{eq:coll_prescr}) and 
(b) erosion, equation (\ref{eq:erosion}). Different colors 
correspond to different size ratios of colliding 
objects indicated on panels. Solid and dashed curves are for 
collisionally strong and weak objects, correspondingly. 
\label{fig:crit_vel}}
\end{figure}

Figure \ref{fig:crit_vel} illustrates the two collisional 
criteria (\ref{eq:coll_prescr}) \& (\ref{eq:erosion}) by showing 
the critical (minimum) relative planetesimal velocity $v_{\rm coll}$ 
that leads to either catastrophic destruction (panel a) or erosion 
(panel b) of the bodies of different sizes. Various curves 
correspond to different size ratios and internal 
strengths of the objects involved in a collision. 

In the case of catastrophic disruption critical $v_{\rm coll}$
is a sensitive function of the size ratio of objects involved. 
Collisions of objects of similar size are clearly more 
destructive than those of planetesimals with very different 
sizes as the former are characterized by lower critical $v_{\rm coll}$. 
For collisionally strong objects (solid curves) we find that 
most destructive collisions (requiring the lowest relative 
speed $\sim 10$ m s$^{-1}$ for destruction of equal mass objects)
involve $\sim 300$ m planetesimals, almost independent 
of the mass ratio. For collisionally weak objects this size 
is $\sim 100$ m and $v_{\rm coll}\sim 1$ m s$^{-1}$ for 
$m_1=m_2$. 

In the case of erosion critical $v_{\rm coll}$ attains minimum 
values roughly at the same sizes. However, the dependence on 
mass ratio is very weak and vanishes in the limit of 
$m_2\ll m_1$. This follows from equation (\ref{eq:QR}) that 
demonstrates that in this limit $Q_{\rm R}\propto m_2$, 
canceling the dependence on $m_2$ in the right hand side of 
the condition (\ref{eq:erosion}). This difference in behaviors 
between the two collisional criteria has important implications 
as we show next.

We note at this point that critical velocity curves shown 
in Figure \ref{fig:crit_vel}b are likely to be not applicable 
for the case of erosion by very small objects. In this limit 
one would expect cratering and mass loss from target to be 
determined by its {\it local} material properties (Housen \& 
Holsapple 2009), rather than global ones as suggested by the 
Stewart \& Leinhardt (2009) prescription. Then the critical 
velocity (in the strength-dominated regime, in the absence 
of ejecta re-accumulation) should become independent of the 
target size as the projectile-to-target size ratio tends to 
zero; this is not what Figure \ref{fig:crit_vel}b shows. To 
avoid this issue in the following we do not explore erosion 
in the limit of very large size ratio of colliding bodies,
see \S \ref{sect:erode_role}.


\subsection{Relative velocities and collision outcomes.}  
\label{sect:vel_coll}

We now couple this understanding of different collisional
outcomes with the dynamical results of Paper I and proceed 
as follows. We compute the relative collision velocity of the two 
objects $v_{\rm{coll}}$ as 
$v_{\rm{coll}}^2=e_{12}^2 v_K^2+2G(m_1+m_2)/(d_1+d_2)$, 
where $d_{1}$, $d_{2}$ are the sizes of planetesimals with masses 
$m_{1}$, $m_{2}$. Note that by using the maximum possible 
approach velocity $e_{12} v_K$ for calculating $v_{\rm{coll}}$ 
we are being conservative, 
since the actual approach speed may be as small as 
$(1/2)e_{12} v_K$, see \S \ref{sect:dyn_sum}. The procedure 
used for calculating relative eccentricity of colliding 
planetesimals $e_{12}$ in both the non-precessing 
and precessing disks is outlined in Appendix \ref{eq:coll_out}. 

Maps of $e_{12}$, $v_{12}=e_{12}v_K$ such as the one presented 
in Figure \ref{fig:coll_outcomes} show that $e_{12}$ is a 
function of $d_{1}$, $d_{2}$, meaning that the same is true for 
$v_{\rm{coll}}$ in our approach. We can then use these maps to 
directly illustrate collision criteria for both strong and 
weak planetesimals. In Figure \ref{fig:coll_outcomes}a the two 
regions inside the white boundaries stretching along the 
$d_1=d_2$ line represent the ``zone of destruction'': 
planetesimals with sizes falling into this region get 
catastrophically destroyed in mutual collisions. The extent 
of such zone in $d_p$ is indicated with a white arrow, 
and the largest and smallest planetesimal sizes that get 
destroyed in collisions are denoted $d_l$ and $d_s$.

In Figure \ref{fig:coll_outcomes}b black contours delineate 
``zones of erosion'': collisions of objects falling within 
the corresponding contour result in mass loss by the larger
planetesimal, hindering growth. The extent of the erosion 
zone is characterized by the dimensionless parameter $\chi$,
which is the {\it smallest target-to-projectile size ratio of 
objects that can get eroded} in a collision for a given 
set of system parameters; 
see Figure \ref{fig:coll_outcomes}b for illustration of this 
definition. The overall morphology of 
the erosion zone is similar to ``erosion regions'' found by 
Th\'ebault \etal (2008) in $d_1-d_2$ space using numerical 
integration of planetesimal orbits and fragmentation 
criteria different from ours, see their Figures 2, 6, 7. 
Note however that our Figure \ref{fig:coll_outcomes}b shows the
erosion zone over much broader range of planetesimal sizes.

Both the ``islands of destruction'' in Figure 
\ref{fig:coll_outcomes}a and the ``islands of erosion'' in 
Figure \ref{fig:coll_outcomes}b exhibit a narrow ``channel'' 
between them at $d_1=d_2$, where the growth is possible. This 
common feature is due to the fact that $e_{12}\to 0$ when 
$d_1$ and $d_2$ are exactly the same, because ${\bf e}_p$ is 
a function of planetesimal size only. At the same time the 
general morphologies of the destruction and erosion regions 
are different --- the former does not extend too far from 
the $d_1=d_2$ line because catastrophic destruction of a
target planetesimal in collision of very different objects 
(either $d_1/d_2\ll 1$ or  $d_1/d_2\gg 1$) would require
very high relative velocity, see Figure \ref{fig:crit_vel}a.
On the other hand, erosion is possible even for collisions of 
highly unequal objects, see Figure \ref{fig:coll_outcomes}b, 
simply because the critical $v_{\rm coll}$ becomes independent 
of $d_1/d_2$ as $d_1/d_2\to 0$. As expected, for collisionally 
weak objects both the destruction and the erosion zones are 
more extended in $d_1-d_2$ space, as shown by the dashed 
contours in Figure \ref{fig:coll_outcomes}. 

\begin{figure}[t]
\vspace{-1cm}
\epsscale{1.2}
\plotone{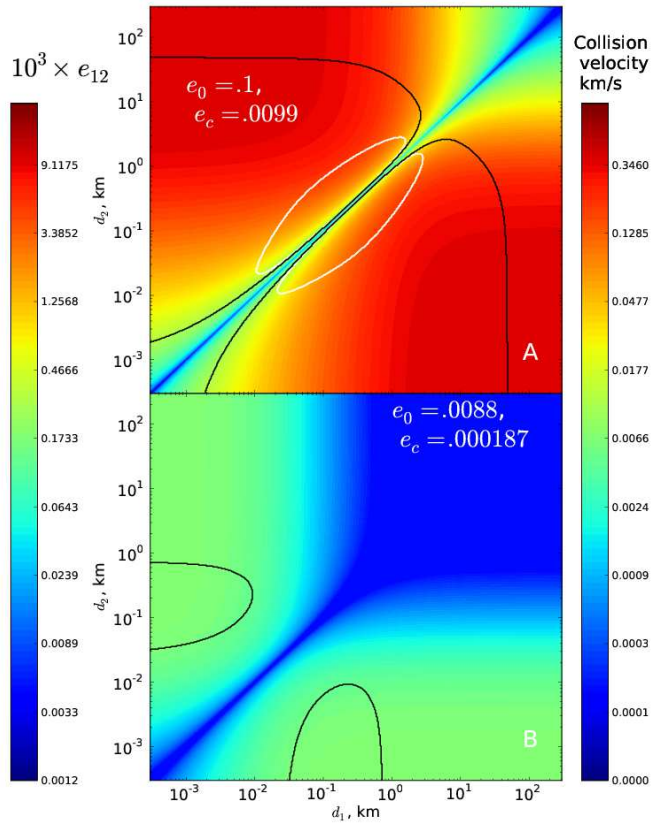}
\caption{
Variation of the destruction (white contours) and erosion 
(black contours) zones with the disk eccentricity and $e_c$. 
Calculations are done for the same parameters as in Figure 
\ref{fig:coll_outcomes}, except that now we use 
(a) $e_0=0.1$ (resulting in $e_c\approx 10^{-2}$) and 
(b) $e_0=8.8\times 10^{-3}$ (resulting in $e_c\approx 
1.9\times 10^{-4}$). Note that in panel (b) catastrophic 
disruption never presents a problem for planetesimal growth
(no white contours).
\label{fig:coll_e_var}}
\end{figure}

The extent of these zones sensitively depends on the value 
of the eccentricity scale $e_c$. This is illustrated in Figure 
\ref{fig:coll_e_var} where the variation of these zones with 
the characteristic planetesimal eccentricity $e_c$ is shown for 
strong planetesimals; the rest of the parameters are as in 
Figure \ref{fig:coll_outcomes}. For high value of the disk 
eccentricity (at its outer edge) $e_0=0.1$ (panel a) one 
obtains high $e_c\approx 10^{-2}$, 
which results in very extended destruction and erosion zones.
The former zone has $d_l/d_s\approx 300$, while for the latter 
$\chi\approx 1$. In other words, the growth-friendly channel
between the two lobes of the erosion zones is extremely narrow, 
making planetesimal agglomeration highly unlikely in this case. 

Lowering $e_0$ to 0.03 ($e_c\approx 2.45\times 10^{-3}$) as in 
Figure \ref{fig:coll_outcomes}a shrinks the size 
of the destruction zone, so that it presents danger for
planetesimals within a size range of only about an order of 
magnitude, $d_l/d_s\sim 10$. Reducing disk eccentricity
even further as in Figure \ref{fig:coll_e_var}b 
($e_0=8.8\times 10^{-3}$, $e_c\approx 1.9\times 10^{-4}$) 
we find the catastrophic destruction zone to fully disappear. 
 
At the same time, erosion zones tend to persist even in 
disks with very small eccentricity. For example, one finds 
$\chi\approx 3$ for $e_0=0.01$ ($e_c\approx 3.15\times 10^{-4}$), 
see Figure \ref{fig:coll_outcomes}b. This means that planetesimals
in such a disk cannot erode a larger object if its mass is 
$\lesssim 30$ times higher. And in Figure \ref{fig:coll_e_var}b, 
where the destruction zone 
vanishes completely, the erosion zone with $\chi\sim 10$
is still present and may affect growth of planetesimals 
with radii $\sim 0.05-1$ km.


\section{Implications for planetesimal growth in binaries.}  
\label{sect:pl_form}


We now use our understanding of the collisional outcomes 
described in \S \ref{sect:coll} to explore the 
possibility of planetesimal growth in binaries as a function 
of the two key protoplanetary disk characteristics --- disk
mass $M_d$ and its eccentricity at the outer edge $e_0$
(defined by equation (\ref{eq:e0}); we fix the disk model to 
have $p=1$, $q=-1$). 

In Figure \ref{fig:aligned_maps} we present maps of collisional 
outcomes for strong planetesimals in the $M_d-e_0$ space. Each 
map uses parameters of a particular planet-hosting binary --- 
HD196885, $\gamma$ Cep, and HD 41004 (Chauvin \etal 2011) --- 
selected because they host Jupiter-mass planets in AU-scale orbits.
These maps are computed at the distance from the primary $a_p$ 
equal to the present-day semi-major axis of the planet (shown
on panels); planet mass is indicated by the vertical red dashed 
line in each panel. Calculations used to produce this figure assume that 
the disk is {\it aligned} with the binary, i.e. $\varpi_d=0$. 
Effect of non-zero $\varpi_d$ is explored further in 
\S \ref{sect:disk_orient}.


\subsection{Accounting for catastrophic disruption.}  
\label{sect:cat_role}

For each point in the two-dimensional space $M_d-e_0$ we construct 
the relative velocity distribution for planetesimals of 
different sizes as shown in Figure \ref{fig:coll_outcomes}.
Using this map of $e_{12}$ and the recipe provided in \S 
\ref{sect:coll} we determine whether the catastrophic destruction 
zone (white contours in Figures \ref{fig:coll_outcomes} \& 
\ref{fig:coll_e_var}) defined by the condition 
(\ref{eq:coll_prescr}) appears 
in it. If it does not, then the corresponding points in 
$M_d-e_0$ space in Figure \ref{fig:aligned_maps} are colored 
grey. The resultant grey region in this Figure covers
part of the parameter space in which catastrophic collisions 
do not present a danger to planetesimal growth. 

In the opposite case, when the white contours appear in the 
$e_{12}$ maps, catastrophic disruption  gets in the way of 
planetesimal growth. Parts of $M_d-e_0$ phase space, in which  
planetesimal growth is interrupted by catastrophic collisions 
are not colored and lie outside the grey regions in Figure 
\ref{fig:aligned_maps}.


\subsection{Accounting for erosion.}  
\label{sect:erode_role}

\begin{figure}
\vspace{-0.9cm}
\epsscale{1.3}
\plotone{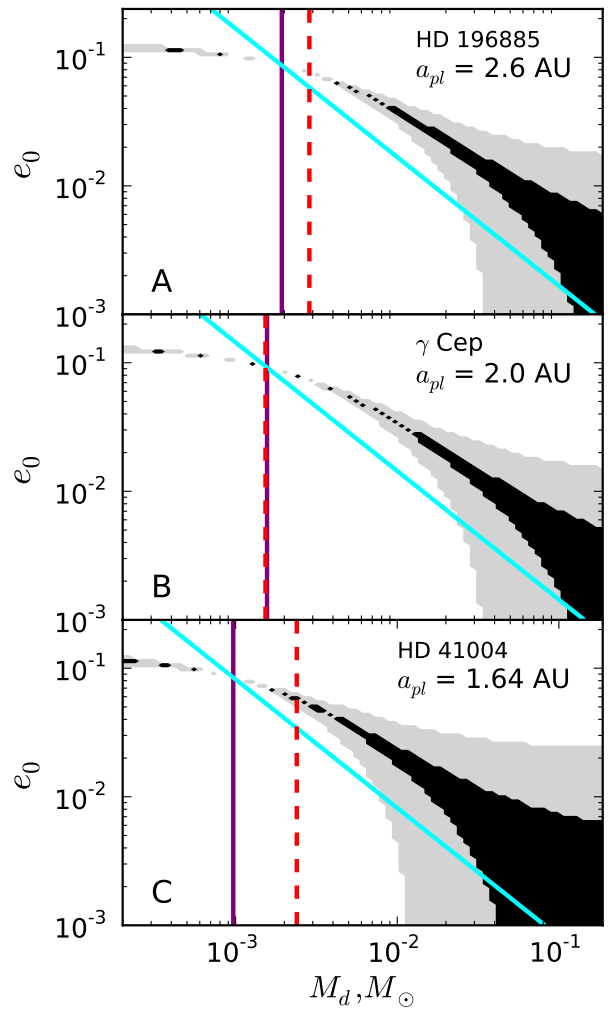}
\caption{
Map of the conditions favorable for planetesimal growth in the 
$M_d-e_0$ space for three binaries (labeled on panels) harboring 
Jupiter mass planets in orbits with $a_{\rm pl}\sim $ AU. 
Planetary semi-major axes are indicated and their $M_{\rm pl}\sin i$ 
are shown with red dashed lines in each panel. Gray areas 
correspond to disk parameters for which the catastrophic 
destruction of planetesimals of any size never happens. Black 
region is a part of phase space where growth with some erosion,
limited by the condition $\chi>\chi_{\rm min}=10^{2/3}\approx 4.6$ 
(see Fig. \ref{fig:coll_outcomes}b) can take place.  The purple 
and cyan lines are the $|A_b|=|A_d|$ and $|B_d|=|B_b|$ 
conditions, i.e. $M_d=M_{d,A=0}$ and 
$M_d=M_{d,|B_d|=|B_b|}$ curves defined by equations (49,PI) 
and (52,PI).
\label{fig:aligned_maps}}
\end{figure}

Even if catastrophic fragmentation is avoided (i.e. outside of 
white region in Figure \ref{fig:aligned_maps}), planetesimal 
growth may still be complicated by the erosion of growing objects 
in numerous collisions with smaller 
planetesimals. To address this issue we check whether for given values 
of $M_d$ and $e_0$ the erosion condition (\ref{eq:erosion}) gets 
satisfied for any $d_1$, $d_2$ in a corresponding map of $e_{12}$ 
(i.e. whether black contours such as in Figures \ref{fig:coll_outcomes} 
\& \ref{fig:coll_e_var} appear in the $e_{12}$ map). If it does,
we need to decide how dangerous it can be for growth,
which is a non-trivial issue.

First, demanding erosion to be {\it completely absent} as a 
necessary condition for planetesimal growth is likely too 
conservative. First, even if some collisions are erosive, 
planetesimals should still be able to grow provided that 
the mass gain in non-erosive collisions exceeds the mass 
loss in erosive impacts.
Examination of the erosion zone shape in Figures 
\ref{fig:coll_outcomes}b \& \ref{fig:coll_e_var} shows that 
a body of a given radius gets eroded predominantly by objects
much smaller in size, which may be chipping off relatively 
small total mass even if erosive collisions are numerous. At the 
same time, collisions with more massive objects of comparable 
size result in mergers, adding substantial amount of mass and
easily resulting in the net mass gain and overall 
growth of planetesimals. This can naturally be the case
if the planetesimal size distribution is such that at all
times most mass is concentrated in largest objects.  

The exact balance of mass loss and gain depends on the 
velocity and mass spectrum of colliding planetesimals. The 
results of Paper I allow us to predict the former. However, 
the latter can be known only after a self-consistent calculation 
of planetesimal coagulation and evolution of the mass spectrum
is performed. Such calculation needs to use the improved 
dynamical inputs from Paper I and requires understanding 
the inclination distribution of planetesimals in binaries, which 
is one of the key inputs for calculation of their collision 
rate. This calculation is beyond the scope of the present work, 
as our present main goal is simply to understand the 
general implications of the improved description of 
planetesimal dynamics (Paper I) on their collisional 
evolution. 

Second, recently Windmark \etal (2012) and Garaud \etal (2013) 
have shown that planetesimal growth can proceed even in the 
presence of collisional barriers. This possibility arises 
when the coagulation-fragmentation process is treated in a 
{\it statistical} sense, allowing for a {\it distribution of 
collisional outcomes}. Unlike the deterministic approach that 
is usually employed, this way of treating planetesimal 
growth allows low probability events --- formation of massive objects 
immune to collisional destruction --- to occur, given a large total 
number of bodies in the system, through a series of ``lucky" collisions. 
As a result, some planetesimals can grow even though the majority 
get destroyed. In our case this may allow growth if 
some degree of erosion and even a chance of catastrophic 
fragmentation (i.e. $d_l/d_s>1$, see Figure 
\ref{fig:coll_outcomes}a) are present.

To account for these arguments we assume planetesimal growth 
to be possible in presence of {\it some erosion}, as long as 
it is not too significant. More specifically, we will assume 
that planetesimals can grow (in a statistical sense) if 
the extent of the erosion zone is limited by some {\it minimum} 
value of the parameter $\chi$ defined in \S \ref{sect:coll} 
and Figure \ref{fig:coll_outcomes}b. In this work, 
we use a fiducial value 
\ba
\chi_{\rm min}=10^{2/3}\approx 4.6, 
\label{eq:chimin}
\ea
which means that 
a growing planetesimal cannot be eroded in collisions with 
projectiles more massive than 
$10^{-2}$ of its own mass. We choose this particular value 
of $\chi_{\rm min}$ simply for illustrative purposes, while in 
practice it should be determined based on planetesimal 
coagulation models (Windmark \etal 2012; Garaud \etal 2013).
It is also low enough that we do not need to worry about the 
applicability of the critical velocity curves in Figure 
\ref{fig:crit_vel}b in the $\chi\to \infty$ limit, see the discussion 
in \S \ref{sect:coll}.  

Black regions in Figure \ref{fig:aligned_maps} cover the 
part of the $M_d-e_0$ parameter space where the condition 
$\chi>\chi_{\rm min}=10^{2/3}$ is fulfilled. We assume 
planetesimal growth to be possible there, despite some degree of
erosion in collisions with small objects.


\subsection{Specific systems.}  
\label{sect:spec_sys}

A general conclusion that can be drawn from Figure 
\ref{fig:aligned_maps} is that, given our growth criteria, 
planetesimal accretion may be possible in tight binaries at 
the semi-major axes of the present day planets, as long as 
the disk mass is high and the disk eccentricity is low. 
Growth is also possible along a narrow extension of the 
colored region towards higher $e_0$ and lower $M_d$, roughly along 
the cyan line $|B_d|=|B_b|$ describing the equality of 
planetesimal excitation by the binary and the disk. The 
origin of this growth-friendly region is connected to the 
existence of the valley of stability (see \S \ref{sect:dyn_sum}) 
in aligned disks, which is further discussed in \S 
\ref{sect:disk_orient}.

Focusing on specific systems, Figure \ref{fig:aligned_maps}c
shows that in situ planetesimal growth (i.e. at the observed 
semi-major axis of the planet) is easiest in the HD 41004
system (Zucker \etal 2004). Planetesimal growth in presence 
of some ($\chi>10^{2/3}$) erosion (black region) is possible 
in this binary even for $M_d\approx 0.02M_\odot$, as 
long as $e_0\lesssim 0.01$. The reason for such favorable 
conditions lies primarily in the relatively small semi-major 
axis of the planet, $a_{\rm pl}\approx 1.6$ AU, on which 
$e_c$ depends very steeply, and the low mass of the primary, 
$M_p\approx 0.7M_\odot$, which lowers $v_K$.

Planetesimal growth is most difficult in HD 196885 system 
(Correia \etal 2008), see Figure \ref{fig:aligned_maps}a. 
Previously, Th\'ebault (2011) realized that HD 196885 
presents the most serious challenge for in situ planetesimal 
growth. This is mainly because of the large 
$a_{\rm pl}\approx 2.6$ AU,
making planetesimal accretion with some erosion possible only 
in very massive disks with $M_d\gtrsim 0.15M_\odot$ and for 
$e_0\lesssim 0.08$. Note that at very high $M_d$ an
{\it evection resonance} corresponding to commensurability 
$A=n_b$ between the planetesimal apsidal precession and 
the binary mean motion (Touma \& Wisdom 1998), can appear 
in the disk. This would additionally disturb dynamics of 
planetesimals and complicate their growth (see Paper I). 

Not too different is $\gamma$ 
Cephei (Figure \ref{fig:aligned_maps}b) with its high 
$a_{\rm pl}\approx 2$ AU and $M_p\approx 1.6M_\odot$: 
here planetesimal growth with $\chi>10^{2/3}$ requires 
$M_d\gtrsim 0.1M_\odot$ and $e_0\lesssim 0.007$. Alternatively,
growth should also be possible if disk parameters fall within the 
{\it valley of stability} (see \S \ref{sect:disk_orient}), 
which can be quite wide at its lower right end.

Figure \ref{fig:aligned_maps} reveals some additional important 
details. First, purple vertical lines in Figure 
\ref{fig:aligned_maps} mark the location of the {\it secular 
resonance}, where the planetesimal precession rate 
$A=A_b+A_d$ becomes zero, see \S 7.1 of Paper I. At the disk 
mass $M_{d,A=0}$ corresponding to this resonance (equation 
(49,PI)) the value of $e_c$ diverges in secular approximation, 
meaning that planetesimals collide at very high speeds 
resulting in their destruction. 
 
\begin{figure}
\epsscale{1.25}
\plotone{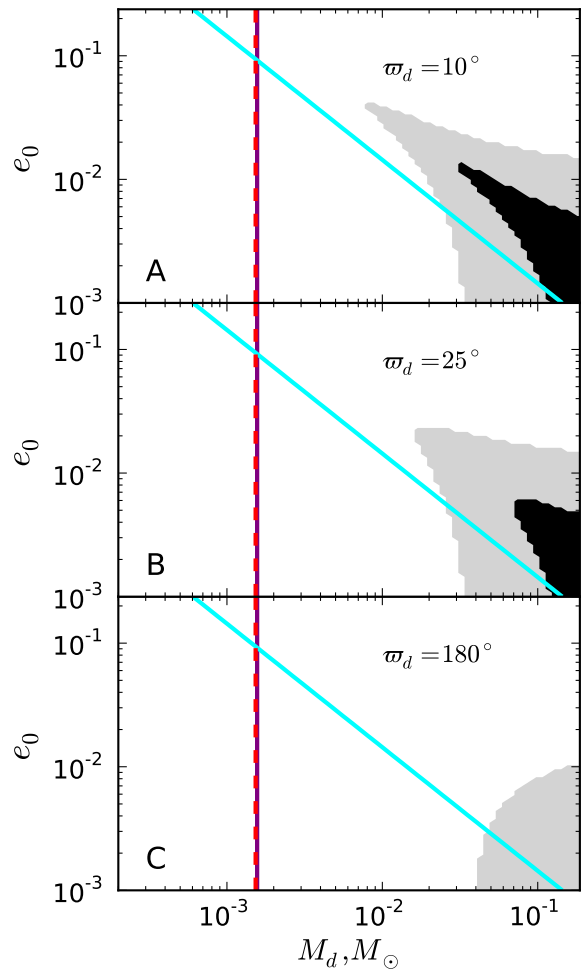}
\caption{
Same as Figure \ref{fig:aligned_maps}b (i.e. growth zones
in $\gamma$ Cep) but for a disk misaligned from the binary 
apsidal axis by an angle $\varpi_d$ indicated on each panel.
Note the gradual disappearance of the ``valley of stability''
and shrinking of the region favorable for planetesimal growth
as $\varpi_d$ is increased.
\label{fig_disk_orient}}
\end{figure}

Second, planet masses ($M_{\rm pl}\sin i$) indicated by 
red dashed lines in Figure \ref{fig:aligned_maps} never 
fall below $M_{d,A=0}$ at the corresponding semi-major axis. 
Under the natural assumption $M_d>M_{\rm pl}$ we can conclude 
that the protoplanetary disk mass $M_d$ must have exceeded 
$M_{d,A=0}$ by at least a factor of several. 
Based on the results of SR13 and Paper I, this inevitably 
implies that the in situ growth of planetesimal 
towards forming cores of gas giants should {\it always proceed 
in either DD or DB dynamical regimes} in the 
classification of SR13, i.e. when $|A_d|\gtrsim |A_b|$ (to the right 
from the purple line in Figure \ref{fig:aligned_maps}) and 
{\it disk gravity dominates planetesimal precession rate}. This 
important fact was completely overlooked prior to the work of 
R13 and SR13.


\section{Sensitivity to model parameters}  
\label{sect:sens}


Next we explore the sensitivity of our results to the different 
parameters of the calculation, such as the disk orientation 
(\S \ref{sect:disk_orient}), radial distribution of the gas 
surface density and eccentricity (\S \ref{sect:disk_model}), 
distance from the primary (\S \ref{sect:disk_ecc}). 
We focus on the $\gamma$ Cephei system and vary our inputs 
one by one. The results are then compared with Figure 
\ref{fig:aligned_maps}b, allowing us to isolate the most 
important factors affecting planetesimal growth.


\subsection{Role of the disk orientation.}  
\label{sect:disk_orient}

We start by analyzing how planetesimal growth is affected 
as we vary the disk orientation with respect to the
binary apsidal line, quantified via the angle $\varpi_d$.

An important feature of the perfectly aligned disk visible in Figure 
\ref{fig:aligned_maps}b is the {\it ``safe zone''} favorable for growth, 
which extends towards the upper left corner of the $M_d-e_0$ map. 
Its origin lies in the presence of the dynamical  
{\it ``valley of stability''} in the the $M_d-e_0$ phase space for
aligned disks. This feature is easily visible in Figure 4a of 
Paper I as a narrow region, within which characteristic eccentricity 
$e_c$ is low. Comparing with Figure \ref{fig:aligned_maps}b we see 
that the shape of the growth-friendly region in $M_d-e_0$ space 
mirrors the overall morphology of the dynamical valley of 
stability. 

An in-depth discussion of the ``valley of stability'' 
properties is provided in \S 7.2 of Paper I, where it is shown, 
in particular that for $M_d\gtrsim M_{d,A=0}$ this valley 
stretches close to $|B_b|=|B_d|$ curve (cyan line in 
Figures \ref{fig:aligned_maps} \& \ref{fig_disk_orient}) 
defined by equation (52,PI), which corresponds to the equality 
of the planetesimal eccentricity excitation contributions 
provided by the disk ($B_d$) and the binary companion ($B_b$).
This dynamical feature makes planetesimal growth possible 
even in low-mass disks with $M_d\gtrsim 3\times 10^{-3}M_\odot$
as long as the disk eccentricity $e_0$ takes on a particular value
of order several per cent.
The valley of stability vanishes for $M_d\sim M_{d,A=0}$ 
($\approx 1.6\times 10^{-3}M_\odot$ for $\gamma$ Cep) because 
a secular resonance appears at this disk mass driving $e_c$ to very 
high values and making growth impossible. However, for even 
lower disk masses the valley of 
stability re-emerges, making planetesimal growth possible 
even in low mass disks ($M_d\lesssim 10^{-3}M_\odot$) but only 
at a certain (narrow) range of the disk eccentricity $e_0\approx 0.1$ 
given by equation (54,PI). 

As the disk orientation changes away from perfect alignment, the 
valley of stability starts to shrink. Figure \ref{fig_disk_orient}a
shows that even relatively small misalignment of $\varpi_d=10^\circ$
is enough to eliminate the growth-friendly zone for 
$M_d\lesssim M_{d,A=0}$. Planetesimal growth without {\it 
catastrophic disruption} is then possible only for 
$M_d\gtrsim 10^{-2}M_\odot$, but it may still proceed at
disk eccentricity $e_0\sim 0.01-0.04$ (upper left of the 
grey region). Growth allowing 
for {\it some erosion} with $\chi <10^{2/3}$ requires 
$M_d\gtrsim 0.03 M_\odot$ and $e_0\lesssim 0.015$ 
(upper left of the black region). 
At the same time the overall morphology of the growth-friendly 
zone remains roughly the same as in the aligned case --- a 
relatively narrow region extending towards the upper left 
corner of the $M_d-e_0$ parameter space.

At $\varpi_d=25^\circ$ growth avoiding the catastrophic 
fragmentation is possible if $M_d\gtrsim 0.02M_\odot$ and 
$e_0\lesssim 0.02$. Erosion with $\chi < 10^{2/3}$ is not 
an obstacle for growth only for $M_d\gtrsim 0.08M_\odot$ and 
$e_0\lesssim 0.006$.  

Finally, for an anti-aligned disk ($\varpi_d=\pi$) the growth 
avoiding catastrophic destruction is still possible for 
$M_d>0.04M_\odot$, $e_0<0.01$. Planetesimal growth with even 
modest erosion ($\chi<10^{2/3}$) is certainly not possible in 
such a disk if its mass is below $\sim 0.2M_\odot$. 

\begin{figure}
\vspace{-0.9cm}
\epsscale{1.25}
\plotone{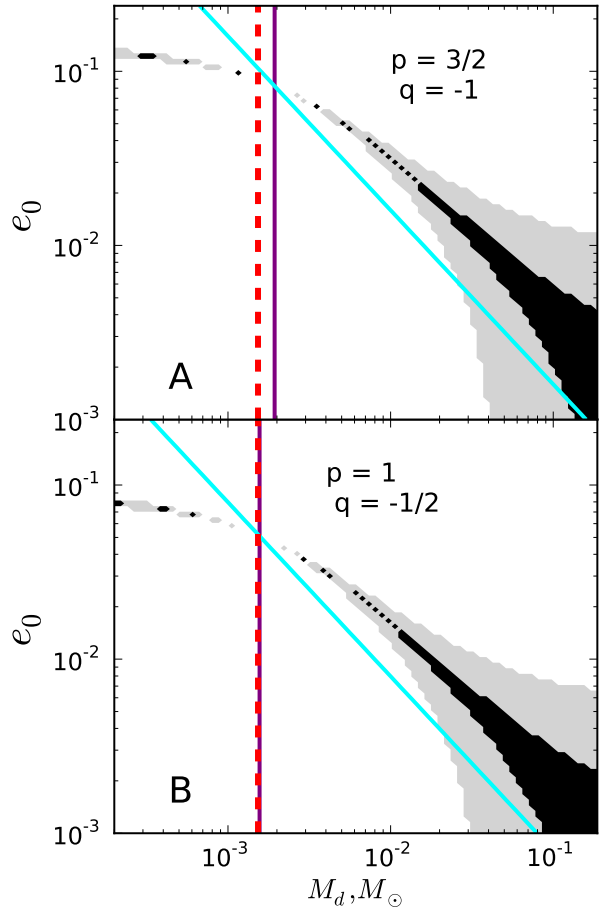}
\caption{
Same as Fig. \ref{fig_disk_orient}a (aligned disk at 2 AU in $\gamma$ 
Cephei) but for two different disk models with parameters indicated on 
panels. For comparison, Fig. \ref{fig_disk_orient}a uses $p=1$, $q=-1$.
\label{fig:model_comparison}}
\end{figure}

These results demonstrate that both the valley of stability and 
the extended region favorable to planetesimal growth in Figure 
\ref{fig_disk_orient} are endemic to relatively well-aligned 
disks. We conclude that the maximum disk misalignment at which
the valley of stability can still facilitate planetesimal growth is 
$\varpi_d\approx 10^\circ-15^\circ$. 

Simulation results regarding the value of $\varpi_d$ for 
non-precessing disks are rather mixed. Most of the simulations of 
M\"uller \& Kley (2012) are consistent with relatively 
well-aligned disks and $\varpi_d<10^\circ$. This would greatly 
facilitate planetesimal growth in binaries. At the same time, 
Paardekooper \etal (2008) and Marzari \etal (2012) find 
$\varpi_d\approx \pi$, i.e. anti-alignment. Part of the reason 
for the discrepancy between the different studies may lie in 
the method used to determine 
disk eccentricity (Marzari \etal 2009) --- whether it is based 
on osculating orbital elements of fluid elements or on fitting 
the isodensity contours of the disk. Thus, the numerical 
evidence regarding the actual value of $\varpi_d$ is inconclusive 
at the moment.


\subsection{Sensitivity to the disk model.}  
\label{sect:disk_model}

In Figure \ref{fig:model_comparison} we test the sensitivity  of
our results on collisional outcomes to other details of the adopted 
disk model. Namely, we vary power law indices $p$ and $q$ 
characterizing $\Sigma_d(r)$ and $e_d(r)$. Comparison with 
the middle panel of Figure \ref{fig:aligned_maps} shows that variations 
of the $\Sigma_d$ profile (i.e. of $p$) do not induce noticeable 
changes. However, results are sensitive to the eccentricity profile 
--- the model with $q=-1/2$ ($e_g\propto a_d^{1/2}$) in Figure 
\ref{fig:model_comparison}b yields higher disk eccentricity 
$e_g$ at the same semi-major axis and {\it for the same} $e_0$ than 
the $q=-1$ model ($e_g\propto a_d$), see equation 
(\ref{eq:e0}). This has detrimental effect on planetesimal growth 
and shrinks the size of the growth-friendly zone in the $M_d-e_0$
space.


\subsection{Variation with the location in the disk.}  
\label{sect:disk_ecc}

\begin{figure}
\vspace{-0.9cm}
\epsscale{1.25}
\plotone{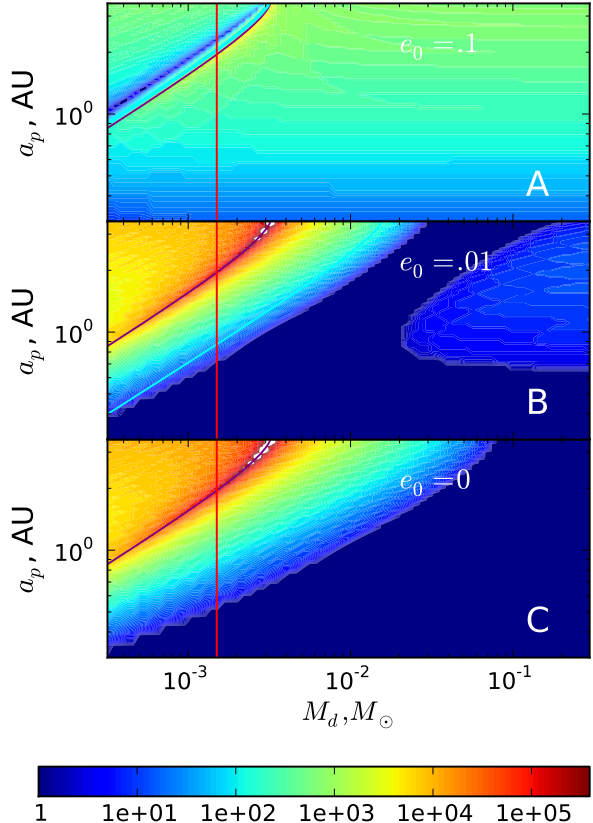}
\caption{
Map of the ratio $d_l/d_s$ (see \S \ref{sect:coll}) in 
the $M_d-a_p$ space, illustrating the possibility of 
catastrophic disruption of planetesimals at different 
locations in the disk. Calculation is done for an aligned 
disk ($\varpi_d=0$) in $\gamma$ Cep system for three different 
values of $e_0$ --- the disk eccentricity at its outer edge, 
indicated on panels. Red line denotes $M_{\rm pl}\sin i$, purple 
and cyan are the $|A_b|=|A_d|$ and $|B_d|=|B_b|$ 
conditions.
\label{fig:map_in_a}}
\end{figure}

\begin{figure}
\vspace{-0.9cm}
\epsscale{1.25}
\plotone{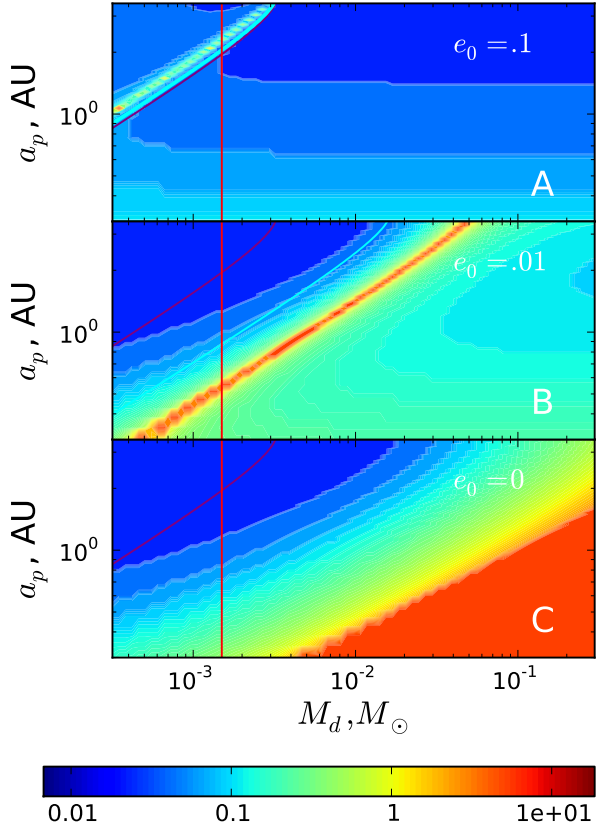}
\caption{
Map of the size ratio $\lg\chi$ (see \S \ref{sect:coll}) in the 
$M_d-a_p$ space, illustrating the sensitivity of planetesimal
growth to erosion. Note that these are logarithmic maps of 
$\lg\chi$ (by definition $\chi>1$, see Figure \ref{fig:coll_outcomes}, 
and the color scheme starts at zero), i.e. yellow corresponds to 
$\chi\approx 10$, when erosion by objects $\approx 10$ times 
smaller than the target size becomes possible for some 
planetesimal sizes. The parameters of the calculation and 
meaning of different curves are the same as in Figure 
\ref{fig:map_in_a}.
\label{fig:map_in_a_er}}
\end{figure}

Calculations shown in Figure \ref{fig:aligned_maps} are performed
at a single location --- present day semi-major axis of the planet 
in each system. In Figures \ref{fig:map_in_a} and 
\ref{fig:map_in_a_er} we illustrate how the conditions
favorable for planetesimal growth change as the distance to the 
star is varied. 

Our discussion of collisional outcomes in \S \ref{sect:coll} shows
that the detrimental effect of catastrophic collisions 
for planetesimal growth can be characterized by the sizes 
$d_l$ and $d_s$ of the largest and smallest objects that 
get destroyed, see Figure \ref{fig:coll_outcomes} for 
illustration. We can describe the effect of catastrophic
collisions via the ratio $d_l/d_s$, which exceeds unity 
whenever such collisions are possible for some planetesimal 
sizes. The higher is $d_l/d_s$, the more extended is the 
catastrophic disruption zone and the more difficult it is
for growing planetesimals to avoid being destroyed in such 
collisions. The white regions in maps in Figure 
\ref{fig:aligned_maps} correspond to $d_l/d_s>1$, while in 
the grey regions catastrophic collisions are absent for 
any planetesimal sizes. 

In Figure \ref{fig:map_in_a} we illustrate the sensitivity 
of planetesimal growth to catastrophic disruption by showing 
the maps of $d_l/d_s$ as a function of both the disk mass 
$M_d$ and the semi-major axis $a_p$, for several values of 
the disk eccentricity at its outer edge $e_0$. Calculation is 
done for an aligned disk in $\gamma$ Cep system.

For a high $e_0 = 0.1$ we see two regions favorable to growth
(i.e. the ones where $d_l/d_s$ is unity or at least less than 
$\sim 10$). First, there is a thin dark blue 
band along the $|B_b|=|B_d|$ (cyan) curve, corresponding to the 
``valley of stability", see equation (52,PI). Second, close 
to the star planetesimal dynamics is completely dominated by 
the disk gravity
(DD dynamical regime in classification of SR13), so that 
$e_c\sim e_g$, which is small in the inner disk (for our $q=-1$).  

For the lower eccentricity models shown in panels (b) and (c), most
of the DD regime (high $M_d$, small $a_p$, see SR13) is favorable 
for planet 
formation. It may seem surprising that the $e_0 = 0.01$ case appears 
to be slightly more favorable than the $e_0 = 0$ case. This is because 
of the existence of the valley of stability for $e_0\neq 0$ (panel b), 
which slightly widens the growth-friendly zone in the DD regime, see
\S 7.2 of Paper I.  

It is also interesting that the upper left corner of the high-$e_0$ 
map shown in panel (a) is more favorable for planetesimal growth than 
in maps corresponding to lower $e_0$. This is caused by the 
degeneracy of the particular choice $e_0=0.1$ mentioned in \S 7.2 
of Paper I (see equation (54,PI)), which causes $e_c$ to be low 
in the corresponding region (BB regime in classification of SR13) 
of Figure 4c of Paper I. 

Next, in Figure \ref{fig:map_in_a} we illustrate the sensitivity 
of planetesimal growth to {\it erosion} by showing the maps of 
$\lg\chi$, where $\chi$ is the lowest target-to-projectile size 
ratio for which erosion is possible for some planetesimal size, 
see \S \S \ref{sect:coll} and Figure \ref{fig:coll_outcomes}
for details. Large values of $\lg\chi$ (red) correspond to the
situation when erosion occurs only in collisions with very small 
objects, which do not result in appreciable mass removal from 
the target. Such collisions are unlikely to prevent planetesimal 
growth as long as such small objects do not account for the 
dominant fraction of the disk mass.

One can see that the behavior of $\lg\chi$ in $M_d-a_p$ space 
largely replicates that of $d_l/d_s$ in Figure 
\ref{fig:map_in_a} --- safe zones near the valley of stability,
as well as at high $M_d$ and small $a_p$. Growth-unfriendly
regions (blue) lie towards higher $a_p$ and at small disk 
masses. Thus, planetesimal growth is easiest in massive disks 
and closer to the star.


\section{Planetesimal growth in precessing disks.}  
\label{sect:pl_prec}

\begin{figure}
\vspace{-0.9cm}
\epsscale{1.25}
\plotone{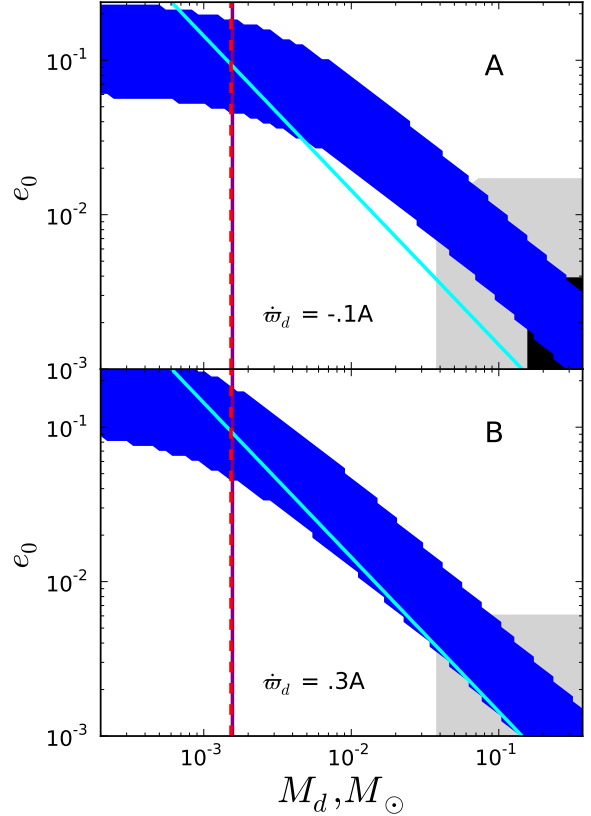}
\caption{
Same as Fig. \ref{fig:aligned_maps}b but for a precessing disk with 
$p=1$, $q=-1$ around $\gamma$ Cephei at 2 AU. Disk precession rate 
is indicated on each panel in units of the local planetesimal 
precession rate $A$ (which itself depends on $M_d$ within each 
panel). Analytical description of planetesimal dynamics fails 
within blue regions, which are excluded from the panels. 
See text for discussion.
\label{fig:prec_disk}}
\end{figure}

In this section we analyze planetesimal growth in disks which do not have 
fixed orientation with respect to the binary orbit but {\it precess} 
at some rate $\dot\varpi_d$. We do this by following the same procedure 
as in \S \ref{sect:coll}, but calculating the relative planetesimal
velocity using the results of \S 6 of Paper I, see Appendix 
\ref{eq:coll_out}. Results are shown in Figure \ref{fig:prec_disk} 
where we display regions in the $M_d-e_0$ space favorable for 
planet formation at 2 AU in $\gamma$ Cephei for two different 
values of the disk precession rate $\dot\varpi_d$, expressed here 
in units of the local value of the planetesimal precession rate 
$A$. Note that the value of $A$ varies within each panel since 
it is a function of $M_d$. 

Calculations described in Appendix \ref{eq:coll_out} for
the case of precessing disk do not provide an analytical solution 
for $\left|\left(A-\dot\varpi_d\right)e_g+B_d\right|\sim |B_b|$
(here, again, $A=A_d+A_b$), which excludes certain  
parts of the $M_d-e_0$ phase space (blue bands) 
from Figure \ref{fig:prec_disk}. In the rest of the figure we 
use the results for strong (\S 6.1 of Paper I) and
weak (\S 6.2 of Paper I) binary perturbation cases,
depending on the circumstances. This makes our treatment of 
collision outcomes in precessing disk somewhat approximate. 
Nevertheless, we can understand the main effects of disk 
precession on planetesimal collisional outcomes by comparing 
these results with Figure \ref{fig:aligned_maps}b. 

First of all, the valley of stability ceases to exist because 
disk-secondary apsidal alignment is not possible in a precessing 
disk. This tends to reduce the size of the growth-friendly zone in
precessing disks, even far from the center of the valley of 
stability.

Second, in the strong binary perturbation regime, below the 
blue band, planetesimal growth conditions are independent of 
$\dot\varpi_d$. This is because the ${\bf e}_p$
solutions obtained in \S 6.1 of Paper I for this regime are 
independent of $\dot\varpi_d$, since eccentricity excitation 
by the disk is weak. The size of the low-$e_0$ growth-friendly 
region varies only because the extent of the excluded 
region (blue band) depends on $\dot\varpi_d$

Third, in the weak binary perturbation regime, above the 
blue band, the extent of the growth-friendly zone does 
depend on $\dot\varpi_d$. To understand this dependence 
we recall (see \S 6.2 of Paper I) that the overall planetesimal
eccentricity scale in a precessing disk is given by $e_c^{\rm pr}$,
defined by equation (45,PI). Above the blue band $A\approx A_d$, 
and we can use expressions (6,PI) and (8,PI) for $A$ and 
$B\approx B_d$.
Recalling that for our disk model with $p=1$, $q=-1$ the 
coefficients in these expressions are 
$\psi_1=-0.5$, $\psi_2=1.5$ we can write
\ba
\frac{e_c^{\rm pr}}{e_g}=1.5
\left(1-\frac{\dot\varpi_d}{A}\right)^{-1}-1.
\label{eq:e_c_pr}
\ea
This ratio is equal to 0.5 in a non-precessing disk,
when $e_c^{\rm pr}\to e_c$. A simple analysis
of equation (\ref{eq:e_c_pr}) then shows that 
$|e_c^{\rm pr}|<e_c$ and non-zero precession {\it suppresses 
planetesimal eccentricity and relative velocity 
compared to the case of a non-precessing disk} if 
\ba
-2<\frac{\dot\varpi_d}{A}<0.
\label{eq:condit}
\ea
Given that for high $M_d$, planetesimal dynamics is in the DD 
regime, for non-pathological disk models (i.e. for 
surface density slope $0<p<3$, SR13) $A<0$, i.e. planetesimal 
apsidal precession is {\it retrograde} relative to 
its mean motion. Then we conclude from the condition 
(\ref{eq:condit}) that {\it slow ($|\dot\varpi_d|<2|A|$) 
prograde precession of the disk is favorable for planetesimal 
growth}. This is indeed seen at the high $M_d$ end of Figure 
\ref{fig:prec_disk}a, although the magnitude of the effect 
is small because of the small adopted value of the 
$|\dot\varpi_d/A|=0.1$ (see below for the characteristic 
value of $|\dot\varpi_d/A|$). 

On the contrary, retrograde or fast prograde ($\dot\varpi_d>2|A|$)
disk precession {\it shrinks} the size of the growth friendly zone,
as demonstrated by Figure \ref{fig:prec_disk}b,c for 
$\dot\varpi_d=0.3A$. This is a bit
counter-intuitive as one may naively expect fast precession 
to result in effective azimuthal averaging of the disk potential,
suppressing planetesimal eccentricity excitation by the 
non-axisymmetric component of the disk gravity, and lowering
$|{\bf e}_p|$ in agreement with R13 and SR13. However, this 
argument loses its validity in presence of gas drag, which 
provides an important contribution
to the value of $e_c^{\rm pr}$. For that reason planetesimal
growth is facilitated by disk precession only when the somewhat 
non-trivial condition (\ref{eq:condit}) is fulfilled.

For our fiducial disk with $p=1$ one finds (R13)
\ba
|A_d| &=& n_p\frac{M_d}{M_p}\frac{a_p}{a_{\rm out}}=
n_b\left[\frac{(1-\mu)a_b^3}{a_pa_{\rm out}^2}\right]^{1/2}
\frac{M_d}{M_p}
\label{eq:A_d1}
\\
&\approx & 
0.1 n_b\frac{a_{b,20}^{3/2}}{a_{\rm out,5}a_{p,2}^{1/2}}
\frac{M_d/M_p}{10^{-2}},
\nonumber
\ea
where the numerical estimate is for the $\gamma$ Cep parameters 
and $n_b=[G(M_p+M_s)/a_b^3]^{1/2}$ is the mean rate of the binary.

At the same time,
simulations of disks in eccentric binaries tend to find a variety of 
outcomes depending on the detailed physics that goes into the 
calculations, with both prograde (Okazaki \etal 2002; Marzari \etal 
2009) and retrograde (Kley \& Nelson 2008; M\"uller \& Kley 2012) 
precession possible. Numerical results suggest that typically 
$|\dot\varpi_d| \sim (1-2)\times 10^{-2}n_b$ (Marzari \etal 
2009; M\"uller \& Kley 2012), which is considerably slower 
than $|A_d|$ evaluated at the semi-major axis of the planet, 
$|\dot\varpi_d|\sim 0.1|A_d|$. In this case, 
according to Figure \ref{fig:prec_disk}a, even if precession 
is prograde its effect on planetesimal growth in high-mass 
disks is going to be small (or slightly negative mainly 
through the elimination of the valley of stability in precessing
disks). 

Lower mass disks ($M_d\sim 10^{-3}M_\odot$), 
containing enough mass to form only terrestrial or 
Neptune-like planets have lower $|A|$. If they precess at
the slow rates found in simulations they may have 
$|\dot\varpi_d|\sim |A|$ satisfied. However, as shown in Figure 
\ref{fig:prec_disk}, planetesimal growth is strongly suppressed
in such low-mass disks. Thus, planetesimal growth in low-mass 
precessing disks must be rather difficult, at least at separations 
$\gtrsim 1$ AU. This is contrary to the non-precessing aligned
disk case, in which the existence of the valley of stability
permits collisional growth even for 
$M_d\lesssim 10^{-2}M_\odot$, see \S \ref{sect:pl_form} and Figure 
\ref{fig:aligned_maps}.

It is also worth noting that simulations with improved treatment 
of the gas thermodynamics (Marzari \etal 2012; M\"uller \& Kley 
2012) and including self-gravity (Marzari \etal 2009) tend to 
produce {\it non-precessing} disks, properties of which we explored 
in previous section.
Thus, disk precession is unlikely to strongly affect our 
conclusions regarding planetesimal growth in S-type binaries.


\section{Radial migration of planetesimals.}  
\label{sect:pl_migr}

Apart from the eccentricity evolution, the non-conservative gas drag 
causes inspiral of planetesimal orbits --- an effect that 
was not accounted for in SR13. We now turn our attention to this 
important issue.

Calculation of the radial drift $\dot a_p$ is 
a more delicate procedure than that of the eccentricity damping. 
As shown by Adachi \etal (1976), even in the case of a circular 
disk one has to account for the radial variation of the gas
density $\rho_g$ when computing $\dot a_p$. 
Calculation becomes even more complicated in the case of an eccentric 
disk with its non-axisymmetric surface density profile. 
Accounting for the difference in azimuthal velocities of
gas and particles that results from the radial pressure gradient 
can be highly non-trivial in the case of an eccentric disk. 

For that reason, we have chosen to describe radial planetesimal
drift $\dot a_p$ using an empirical generalization of the 
appropriate results of Adachi \etal (1976) for the case of an 
eccentric disk. This generalization is physically motivated and 
reduces to the known results in the case of the circular disk
with $e_g=0$. Namely, we use
the equation (4.21) of Adachi \etal (1976), in which we simply set 
$i=0$ and replace $e$ with the relative particle-gas eccentricity
$e_r$. As a result, we find
\ba
\dot a_p=-\pi\frac{a_p}{\tau_a}
\left(\frac{5}{8}e_r^2+\eta^2\right)^{1/2}
\left[\left(\frac{\alpha}{4}+\frac{5}{16}\right)e_r^2+\eta\right],
\label{eq:rad_drift}
\ea
where 
\ba
\tau_a & = & e_r\tau_d=\frac{4\pi}
{3C_D\mbox{E}\left(\sqrt{3}/2\right)}D^{-1}
\label{eq:tau_a}\\
& \approx & 6~\mbox{yr}~C_D^{-1}
\frac{a_{\rm out,5}a_{p,1}}{M_{p,1}^{1/2}M_{d,-2}}
\frac{h/r}{0.1}d_{p,1}.
\nonumber
\ea
is the characteristic timescale, $\tau_d$ is the eccentricity 
damping time defined by equation (18,PI), and 
\ba
\eta=\frac{1}{2}\left(\alpha+s\right)
\left(\frac{c_s}{nr}\right)^2=\frac{1}{2}
\left(p+\frac{s+3}{2}\right)\left(\frac{h}{r}\right)^2
\label{eq:eta}
\ea
is the measure of the azimuthal particle-gas drift caused 
by the pressure support in the gas disk. The different parameters
entering these expressions are the logarithmic slopes of the gas
density and temperature $\alpha\equiv -\partial\ln\rho_g/\partial \ln r$ 
and $s\equiv -\partial\ln T_g/\partial \ln r$, related via 
$\alpha=p+(3-s)/2$, see equation (\ref{eq:sig_0}).
 
In this work we will use power-law temperature profile 
$T(r)=T_1\left(r/{\rm AU}\right)^{-s}$, with $T_1$ being the gas
temperature at 1 AU, so that
\ba
\frac{h}{r}\approx 4\times 10^{-2} 
\left(\frac{M_\odot}{M_{p}}\frac{T_1}{400~\mbox{K}}\right)^{1/2}
\left(\frac{r}{\rm AU}\right)^{(1-s)/2}.
\label{eq:h/r}
\ea
In our calculations we normally take $s=1/2$ and $T_1=400$ K 
(the central stars of compact planet-hosting binaries are usually 
somewhat more massive than the Sun).

Note that the characteristic timescale of the radial drift in the 
case $e_r\gg \eta^{1/2}$ is $|d\ln a_p/dt|^{-1}\sim \tau_a e_r^{-3}=
\tau_d e_r^{-2}$, which is much longer than the eccentricity damping 
time $\tau_d$. For smaller $e_r$ migration time lengthens even 
further. The slowness of the radial drift allows us to
treat $a_p$ as a constant while following the evolution of 
planetesimal eccentricities. 

Radial drift depends steeply on $e_r$ and can be rather fast for
strongly dynamically excited planetesimals. Because of the radial 
pressure support in the gaseous disk resulting in the non-zero 
value of $\eta$, $\dot a_p$ does not completely vanish even as 
$e_r\to 0$. This is not the case for eccentricity evolution --- 
eccentricity damping naturally vanishes for $e_r=0$.

\begin{figure}
\vspace{-0.9cm}
\epsscale{1.3}
\plotone{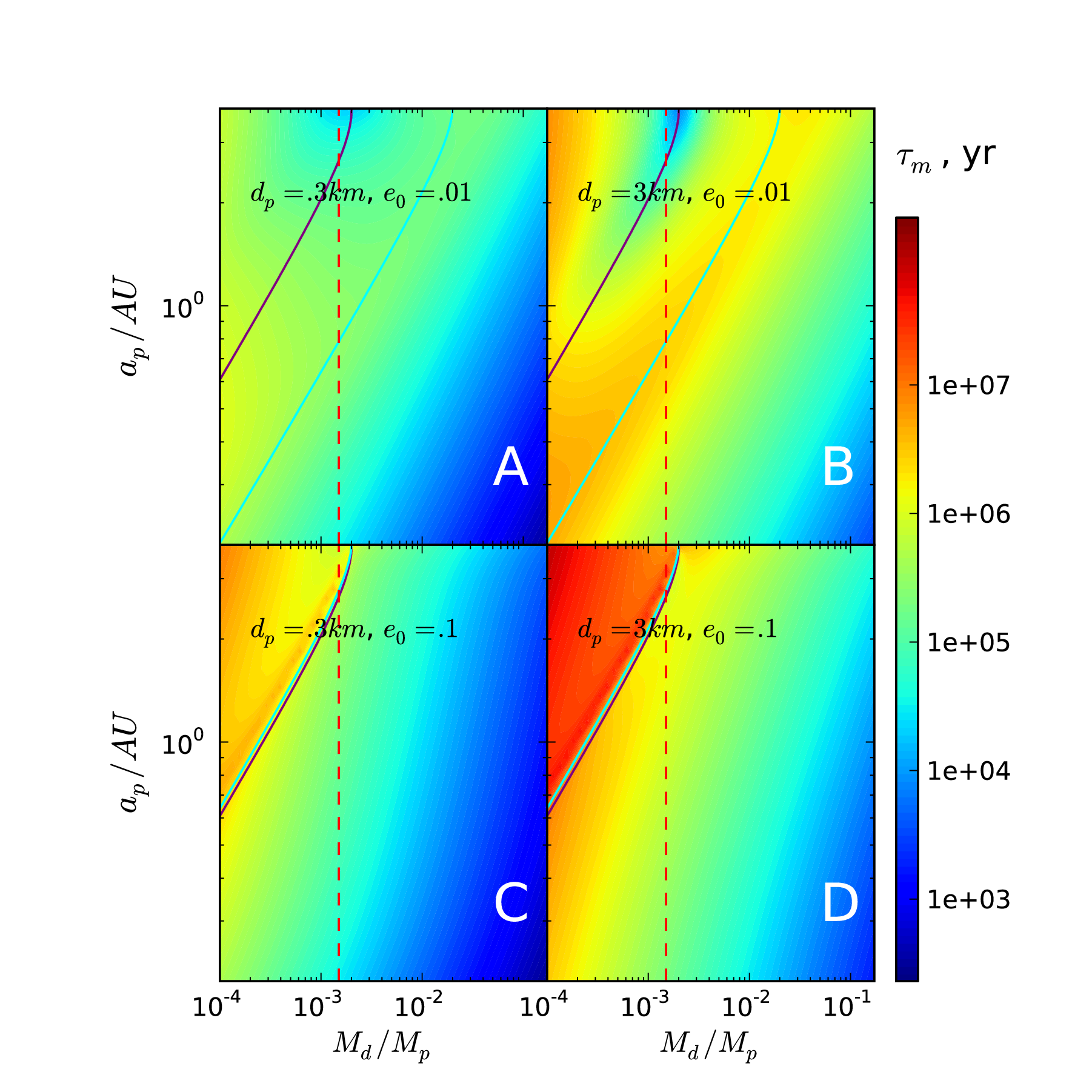}
\caption{
Map of the radial drift timescale $|d\ln a_p/dt|^{-1}$ in 
$M_d-a_p$ space for two different planetesimal sizes, 
$d_p=0.3$ km (left) and $d_p=3$ km (right), and two values of
the disk eccentricity at $a_{\rm out}$, $e_0=0.01$ (top)
and $e_0=0.1$ (bottom). Calculation is done for an aligned 
disk in $\gamma$ Cephei system. 
\label{fig:t_mig}}
\end{figure}

In Figure \ref{fig:t_mig} we map the migration time 
$\tau_m\equiv |a_p/\dot a_p|$ in $M_d$-$a_p$ coordinates.
We calculate $\tau_m$ using equations 
(\ref{eq:rad_drift})-(\ref{eq:h/r}) for our standard (aligned) 
disk parameters in $\gamma$ Cephei for two different values of 
the disk eccentricity and planetesimal size. 

These maps clearly show many non-trivial features and 
significant variation as we change $e_0$ and $d_p$. To better 
understand them we overplot the lines of $A=0$ (purple) and 
$|B_b|=|B_d|$ (cyan) conditions. Interestingly, no significant
feature is seen in the $\tau_m$ maps at the location of the $A=0$ 
secular resonance. This is in contrast to the characteristic 
eccentricity maps in Figure 4 of Paper I, which show the   
divergence of $e_c$ at this resonance caused by 
$e_c\propto |A|^{-1}$ scaling, see equation (29,PI). This 
difference is easily explained by looking at the equation (28,PI),
which shows that the relative planetesimal-gas velocity 
$e_r\propto |A|e_c$ thus removing singularity at $A=0$.
Upon closer inspection one can see only a mild reduction of $\tau_m$ 
in a broad region surrounding $A=0$ curve. It is caused 
by the local $e_r\propto [1+A^2\tau_d^2]^{-1/2}$ dependence on $A$,
increasing $e_r$ and {\it decreasing} $\tau_m$ where $A\to 0$ 
according to equation (\ref{eq:rad_drift}). 

At the same time, in all panels one can easily see a band 
of {\it increased} $\tau_m$, which runs close to the 
$|B_b|=|B_d|$ (blue) curve. Its location is independent of $d_p$ but
is sensitive to $e_0$, with higher disk eccentricity pushing 
this valley of high $\tau_m$ further from the star. Comparing
with Figure 4 of Paper I we conclude that this feature 
is caused by $e_c\to 0$ within this band. Since this is 
possible only in the aligned disk (see \S \ref{sect:disk_orient} 
and \S 7.2 of Paper I) such a feature would not be present in 
a misaligned or precessing disk. 

But in a disk with $\varpi_d\approx 0$, migration time can 
become very long in this region of parameter space: 
$\tau_m\sim$ Myr is quite typical within the valley of 
high $\tau_m$ stretching along the 
$|B_b|=|B_d|$ curve, especially 
for large $d_p$ and higher $e_0$. In this region $e_r$ 
can be so small that $\tau_m$ becomes determined solely 
by the non-zero value of $\eta$ in equation 
(\ref{eq:rad_drift}), which is due to the radial pressure 
support in a gas disk:  
\ba
\tau_m\to\frac{\tau_a}{\pi \eta^2}\propto a_p^{7/4}M_d^{-1}. 
\label{eq:tau_m_sat}
\ea
To arrive at the last scaling we used equations (\ref{eq:sig_0}), 
(\ref{eq:tau_a})-(\ref{eq:h/r}) and adopted $p=1$, $q=-1$.

Figure  \ref{fig:t_mig} shows that $\tau_m$ is higher for
higher $e_0$ in high-$\tau_m$ regions. This is somewhat 
counter-intuitive as one naively expects higher disk 
eccentricity to result in larger planetesimal 
velocities, driving faster, rather than slower, migration. 
This contradiction is resolved by understanding that even 
for the same $d_p$ we are comparing the values of $\tau_m$ 
at special locations, where $e_r\to 0$. Their position 
is roughly described by equation (52,PI) for 
$|B_b|=|B_d|$, from which one infers their 
$a_p\propto (e_0 M_d)^{1/2}$. Plugging
this into equation (\ref{eq:tau_m_sat}) one finds that 
$\tau_m\propto e_0^{7/8}M_d^{-1/8}$, i.e. maximum $\tau_m$
is indeed {\it longer} for higher disk eccentricity. This
is simply a reflection of the fact that for higher $e_0$ the
valley of small $e_r$ moves out to larger $a_p$.
The same reasoning also explains why $\tau_m$ increases 
along the high-$\tau_m$ valley as both $a_p$ and $M_d$ 
get smaller. 

Note the long values of $\tau_m$ in the upper left corner of
Figure \ref{fig:t_mig}c,d. They are caused by a particular
choice of $e_0=0.1$ for which $e_c$ becomes very small 
{\it globally} in the BB regime, when the gravity of the binary companion
dominates over that of the disk (SR13). This coincidence has 
been previously discussed in \S 7.2 of Paper I, see equation 
(54,PI).

Existence of a localized peak of $\tau_m$ has important implications
for planetesimal growth. In a disk with fixed values of $e_0$
and $M_d$, planetesimals in the outer parts of the disk 
migrate inward until they reach the high-$\tau_m$ valley. In a 
narrow range of semi-major axes corresponding to this valley 
their radial drift significantly slows down resulting in the
local increase of the surface density of solids of different 
sizes. Given the dramatic local increase of $\tau_m$ one 
can expect planetesimal density there to exceed its initial 
local value by orders of magnitude. 
Moreover, according to Figure 4 of Paper I, the high-$\tau_m$ 
valley is also the location where $e_c$ becomes very small 
providing favorable conditions for planetesimal growth. These 
points are further discussed in \S \ref{sect:bin_pl_form}.


\section{Planet formation in binaries.}  
\label{sect:bin_pl_form}

Now we apply our understanding of planetesimal growth and 
migration described in previous sections to clarify the 
circumstances under which planets of different masses can 
form in disks within binaries.


\subsection{Conditions for giant planet formation.}  
\label{sect:giant_good}

Presence of planets with $M_{\rm pl}\sin i$ of order several
$M_J$ inevitably implies that their parent protoplanetary disks
must have been massive, $M_d\gtrsim 10^{-2}M_\odot$: disk mass 
cannot be much lower than at least several $M_J$,
otherwise disk simply would not contain enough gas to form these 
massive objects. This argument must hold even despite the 
observational evidence against massive disks in 
small separation binaries coming from sub-mm observations 
(Harris \etal 2012). 

Planet masses 
indicated by the vertical red lines in Figure \ref{fig:aligned_maps}
are no more than an order of magnitude lower than $M_d$ at the edge 
of the (grey) growth-friendly zone for $e_0\sim 10^{-2}$.
Then, under the natural constraint $M_d\gtrsim 10^{-2}M_\odot$, 
Figure \ref{fig:aligned_maps} clearly implies that unimpeded 
planetesimal growth leading to giant planet formation at 
AU-scale separations in binaries is {\it possible provided that 
disk eccentricity is low}, $e_0\lesssim 10^{-2}$. 
This is an important requirement for giant planet 
formation in small separation ($a_b\approx 20$ AU) binaries,
which is inspired by planetesimal dynamics alone. It represents 
one of the key results of this work.   

Unfortunately, we do not have direct measurements of circumstellar 
disk eccentricities in young stellar binaries and cannot address the $e_d$ 
constraint directly. Simulations of disks in eccentric binaries 
with $e_b=0.4$ tend to find rather
low values of $e_d\lesssim 0.05$ (Marzari \etal 2009, 
2012; M\"uller \& Kley 2012; Picogna \& Marzari 2013). 
In fact, Reg\'aly \etal (2013) claim that for $e_b>0.2$ 
protoplanetary disk does not develop permanent eccentricity 
in their simulations and deviations from axisymmetry are 
minimal. This is in contrast to simulations of disks in circular 
(or low-$e_b$) binaries, which often demonstrate high 
$e_d\approx 0.3-0.5$
(Kley \etal 2008; Reg\'aly \etal 2011). Such dychotomy is likely
caused by the smaller truncation radii of the disks in high-$e_b$ 
binaries (Reg\'aly \etal 2011) reducing companion perturbation
on them. Disks in circular binaries can extend further out, 
potentially creating conditions for the disk eccentricity 
excitation via the Lubow (1991) mechanism.

Based on this we conclude that the existing numerical results  
{\it are roughly compatible} with the conditions needed 
for overcoming the fragmentation barrier and forming giant 
planets within massive disks 
($M_d\gtrsim 10^{-2}M_\odot$) in AU-scale orbits, namely, low 
$e_0$ of order several per cent, see Figure \ref{fig:aligned_maps}.
Note that in very massive disks ($M_d\gtrsim 0.1M_\odot$) this 
conclusion holds for arbitrary disk orientation 
as well as in precessing disks, see Figures \ref{fig_disk_orient} 
and \ref{fig:prec_disk}. 

Even in high-$M_d$ disks presence of the 
valley of stability facilitates planet formation. 
Figure \ref{fig:t_mig}a,b shows that in low-$e_0$, high-$M_d$
systems the region of long migration time $\tau_m$ corresponds
to semi-major axis of $2-3$ AU. This means that planetesimals 
would {\it preferentially accumulate} at these locations in 
massive disks by gas-driven radial migration. Corresponding 
increase of the surface density of solids, combined with 
the lowered relative velocities of planetesimals at the same 
locations (see Figure \ref{fig:aligned_maps}), could 
considerably facilitate growth of planetary cores. 

Interestingly, three out of five presently known 
planet-hosting tight binaries have planets at 
$a_{\rm pl}=1.6-2.6$ AU, and all three are massive giants
with $M_{\rm pl}\sin i> 1.6M_J$ (Chauvin \etal 2011). We 
suggest that this may be not a coincidence but, possibly, the 
evidence for in-situ formation of these giants, facilitated 
by the local pile up of solids, in low-eccentricity 
($e_0\lesssim 0.01$), high mass 
($M_d\gtrsim 10^{-2}M_\odot$) disks, which were 
aligned ($\varpi_d\approx 0$) with the orbits of their 
binary companions. 

We also speculate that the observed clustering of the binary
eccentricity in $\gamma$ Cep-like systems (with $a_b\approx 20$ 
AU) around $e_b\sim 0.4-0.5$ (Chauvin \etal 2011; Dumusque 
\etal 2012) is directly linked to lower disk eccentricities 
$e_g$ in them, as suggested by simulations (Reg\'aly \etal 2011).
These makes such eccentric binaries {\it more favorable} for 
overcoming fragmentation barrier and forming planets than 
their circular counterparts. And in highly eccentric systems, 
$e_b\to 1$, disks would be truncated at the radii too small 
to contain enough mass for planet formation. Thus, the 
apparent clustering of $e_b$ of compact ($a_b\approx 20$ 
AU) planet-hosting binaries around $0.4-0.5$ may be not 
coincidental.


\subsection{Earth- and Neptune-like planet formation.}  
\label{sect:Earth}

Formation of terrestrial (like in $\alpha$ Cen system, 
Dumusque \etal 2012) or Neptune-size planets may also proceed 
in massive disks, in which case the conclusions of \S 
\ref{sect:giant_good} would apply directly. At the same time, just
based on the mass budget, low-mass planets might also be expected
to form in lower mass ($M_d\sim 10^{-3}M_\odot$) disks. Sub-mm 
observations suggest that such disks are more abundant than their 
more massive counterparts in binaries with separations of order 
several tens of AU  (Harris \etal 2012). However, satisfying 
the planetesimal growth constraints formulated 
in \S \ref{sect:giant_good} for low $M_d$ becomes problematic,
as can be inferred from the presence of extended growth-unfriendly 
(white) zones at small $M_d$ in Figure \ref{fig:aligned_maps}.
According to Figures \ref{fig_disk_orient} and \ref{fig:prec_disk}  
planetesimal growth is essentially impossible in low-$M_d$ disks 
which are mis-aligned with the binary orbit or precess.

However, in aligned disks low-mass planet formation may still 
be possible even for $M_d\lesssim 10^{-2}M_\odot$. In such disks 
the valley of stability (see Figure 
\ref{fig:aligned_maps}) provides the conditions 
favorable for planet formation even for $M_d\lesssim 10^{-2}M_\odot$ 
and for relatively high $e_0\sim 0.1$. 
Moreover, disk evolution may naturally drive even high-$M_d$
systems towards the valley of stability at a given semi-major 
axis. Indeed, even if the disk starts at relatively high
$e_0\sim 0.1$ and high $M_d\gtrsim 5\times 10^{-3}M_\odot$, 
above the black region in Figure \ref{fig:aligned_maps}, over 
time its viscous evolutions will reduce $M_d$ and ultimately 
bring the disk into the valley of stability, making low-mass 
planet formation quite natural at this point. 

Within the localized regions corresponding to the valley of 
stability one would again have a combination of both the increased 
density of solids due to planetesimal accumulation 
induced by the non-uniform planetesimal drift and the
suppression of relative planetesimal velocities. Both 
factors promote planetesimal 
growth. Figures \ref{fig:map_in_a} \& \ref{fig:t_mig} clearly
show that in low mass disks $M_d\gtrsim 10^{-3}M_\odot$ with
relatively high eccentricities $e_0\sim 0.1$ such low-$e_c$ and
high-$\tau_m$ regions lie at semi-major axes of 1-2 AU.
Earth or Neptune-like planets may form there. 

Finally, unimpeded planetesimal growth within 
relatively low mass disks, 
$M_d\lesssim 10^{-3}M_\odot$, may also be possible close to the star,
at sub-AU separations, provided that the disk has low 
eccentricity, $e_0\lesssim 10^{-2}$. This is seen 
in Figure \ref{fig:map_in_a} \& \ref{fig:map_in_a_er}, which 
demonstrate small $d_l/d_s$ and relatively large $\lg\chi$ 
at small $a_p$. Such mode of planet formation may have been 
responsible for the origin of the Earth-mass planet in $\alpha$ 
Cen B (Dumusque \etal 2012).


\subsection{Comparison with previous studies.}  
\label{sect:compare}

Our finding that fragmentation barrier can be overcome, 
opening a way to planet formation at separations
of several AU in tight binaries such as $\gamma$ Cep 
and $\alpha$ Cen is opposite to the conclusions 
of many previous studies 
(Th\'ebault \etal 2008,2009; Th\'ebault 2011).
The main reason for this difference is in the role 
of (generally non-axisymmetric) protoplanetary disk 
gravity, which we account for in secular approximation, 
while other studies included only gas drag 
and perturbations from the companion. As we showed
in Paper I and in this work this aspect really makes 
a big difference for the outcome --- in disks massive
enough to form giant planets, planetesimal precession 
and eccentricity excitation become dominated by the gravity of
the disk rather than of the companion. Thus, it is very 
important that future studies of planet formation in 
binaries, including those that self-consistently evolve 
the disk using direct hydrodynamical simulations, 
account for the gravitational effect of the disk on
planetesimal motion. This has been previously done in 
Kley \& Nelson (2007) and Fragner \etal (2011) but the 
complexity of planetesimal dynamics including disk 
gravity has not been explored in sufficient detail in 
these studies.

On the other hand, some other previous studies have found 
planetesimal growth in tight binaries to be possible. 
Marzari \& Scholl (2000) arrived at this conclusion by 
noticing the apsidal phasing of planetesimal orbits by 
gas drag. But later Th\'ebault \etal (2008) showed the 
associated reduction of the relative speed $v_{12}$ to be a 
consequence of a single planetesimal 
size approximation. Th\'ebault \etal (2006) find growth 
possible for almost circular binaries with small $e_b$,
since in this case eccentricity forcing by the companion 
vanishes. However, simulations show that disks tend to 
develop large eccentricities ($\gtrsim 0.1$) in systems 
with low $e_b$ (e.g. Marzari \etal 2009,2012; Reg\'aly \etal 2011), 
which, with disk gravity included, would have likely 
resulted in severe difficulty of forming planets.


\section{Summary.}  
\label{sect:summ}


We explored planetesimal growth in AU-scale orbits within 
small-separation ($a_b\approx 20$ AU) binaries using a newly developed 
secular description of planetesimal dynamics (Paper I), 
which includes a number of important physical 
ingredients relevant for this problem --- perturbations due 
to the companion, gas drag, and, most crucially, gravitational 
effects of an eccentric disk. We used our results to assess the 
possibility of planet formation in binaries and arrived 
at the following conclusions.

\begin{itemize}

\item 
By exploring outcomes of pair-wise planetesimal collisions we 
identified ranges of planetesimal sizes for which growth by 
coagulation is suppressed (\S \ref{sect:coll}). Inclusion 
of disk gravity is very important for properly determining 
the extent of accretion-unfriendly zones.

\item 
Planetesimal growth uninhibited by fragmentation is possible
for a broader range of parameters ($M_d$ and $e_0$) in disks, 
which are {\it apsidally aligned} with the binary orbit (\S 
\ref{sect:disk_orient}).

\item 
Radial drift of planetesimals caused by gas drag is highly 
non-uniform in aligned disks, with the drift timescale sharply 
peaking at AU-scale separations. This causes accumulation of
planetesimals at the location where their dynamical excitation 
is weak and provides favorable conditions for their growth
(\S \ref {sect:pl_migr}).

\item 
Formation of giant planets in observed (AU-scale) configurations in
eccentric binaries like $\gamma$ Cep is possible in massive and not very 
eccentric disks, $M_d\gtrsim 10^{-2}M_\odot$ and $e_0\lesssim 0.01$ 
(\S \ref{sect:giant_good}). The former condition is consistent 
with the very existence of massive (several $M_{\rm J}$) planets 
in these systems. The latter is in rough agreement with the results of 
simulations, revealing low disk eccentricity in eccentric 
($e_b\approx 0.4$) binaries. Planet formation may be inhibited 
in circular binaries as simulations show disks  
to develop high eccentricity in such systems.

\item
Terrestrial and Neptune-like planets can form in massive 
disks just as giant planets can. Their genesis in the low-mass 
($M_d\lesssim 10^{-2}M_\odot$) disks is possible close to the star 
($a_p\lesssim$ AU) but is generally suppressed further out, at 
$a_p\gtrsim$ AU. However, if the disk and binary periapses are 
aligned, low mass planets can also form in low-$M_d$ disks at 
certain locations (even at $a_p\sim$ AU) where the radially migrating 
planetesimals (1) accumulate and (2) have low relative velocities, 
promoting their growth in mutual collisions.

\end{itemize}

Our results provide a natural way of explaining the existence of
planets in small separation binaries, such as $\gamma$ Cep and 
$\alpha$ Cen, via the improved understanding of planetesimal 
dynamics. This may eliminate the need to invoke more exotic
scenarios for forming such systems. 

Our calculations assessed 
the possibility of planetesimal growth by exploring just the two 
possible collision outcomes --- catastrophic disruption and 
erosion by objects of certain sizes. The full understanding of 
planetesimal growth in binaries will require a self-consistent 
coagulation simulation that would evolve the mass spectrum of 
objects fully accounting for the complexity of their 
dynamics in binaries. 

Methods developed in this 
work will be used to understand formation of
planets in {\it circumbinary} configurations.


\appendix


\section{Relative eccentricities of planetesimals}
\label{eq:coll_out}


To determine the outcome (destruction or no destruction) of 
a collision between two bodies of size $d_1$ and $d_2$ we 
need to calculate their relative eccentricity 
$e_{12}=\left[(h_1 - h_2)^2 + (k_1 - k_2)^2\right]^{1/2}$. 
In the case of a non-precessing disk we do this by first computing  
$A \tau_d$ in terms of $d_p$ and $d_c$ for each planetesimal using 
equation (32,PI), and then plugging it in the equation 
(64,PI) to find $e_{12}$.

For the precessing disk (see \S \ref{sect:pl_prec}) we do not have 
analytical expressions 
for $h_p$ and $k_p$ in general, but we calculate them for two limiting 
cases (strong and weak binary perturbation cases) using the approach 
described in \S 6.1 and 6.2 of Paper I 
correspondingly. We start by evaluating equation (42,PI).  
If $|(A - \dot \varpi_d)e_g + B_d|$ is within a factor of 2 of $|B_b|$, 
we exclude this point of the phase space from our calculation as we do 
not expect analytical limiting behaviors to apply there.  
If $|B_b| > 2|(A - \dot \varpi_d)e_g + B_d|$, then we use equations 
(43,PI) to determine $h_p$ and $k_p$.  If 
$|B_b| < 0.5|(A - \dot \varpi_d)e_g + B_d|$, then we first 
compute $\left(A-\dot\varpi_d\right)\tau_d$ using equation 
(46,PI) and then calculate ${\bf e}_p\approx {\bf e}_{f,d}$ 
via equation (B2,PI) with $B_d=0$ for each planetesimal. 
Even though $k_p$ and $h_p$ are not constant for a given object 
(eccentricity vectors precess together with the disk), their 
difference is constant and is given by 
\ba
e_{12}^2  =  e_1^2+e_2^2-2e_1 e_2\cos(\phi_1-\phi_2),~~~~~~~~~~~~~~~
e_i =  \left[\frac{e_g^2+\tau_d^2(d_i) B_d^2}{1+\tau_d^2(d_i)
\left(A-\dot\varpi_d\right)^2}\right]^{1/2},
\label{eq:e_rel_coll}
\ea
where $e_i$ ($i=1,2$) are the individual forced 
eccentricities for planetesimals of size $d_i$ and $\phi_i$ are 
their apsidal phases (with respect to the instantaneous direction 
of the disk periastron) given by equation (B2,PI)) with 
$\tau_d=\tau_d(d_i)$.

\end{document}